\documentclass[english]{report}
\usepackage{ae,aecompl}
\usepackage[T1]{fontenc}
\usepackage[latin9]{inputenc}
\usepackage[a4paper]{geometry}
\usepackage{color}
\usepackage{float}
\usepackage{url}
\usepackage{amsmath}
\usepackage{graphicx}
\usepackage{setspace}
\usepackage{amssymb}
\usepackage{nomencl}
\providecommand{\makenomenclature}{\makeglossary}
\makenomenclature

\newcommand{\noun}[1]{\textsc{#1}}
\providecommand{\tabularnewline}{\\}
\usepackage{babel}

\begin{document}

\title{\noindent \textbf{\textsc{Simulations of Doppler Effects in Nuclear
Reactions for AGATA Commissioning Experiments }}}

\author{A\noun{\small li} A\noun{\small l}-A\noun{\small dili}\\
Nuclear Structure Group\\
Division of Nuclear and Particle Physics\\
Department of Physics and Astronomy, Uppsala University\\
UPPSALA, SWEDEN\\
2009\\
}

\date{\includegraphics[clip,scale=0.5]{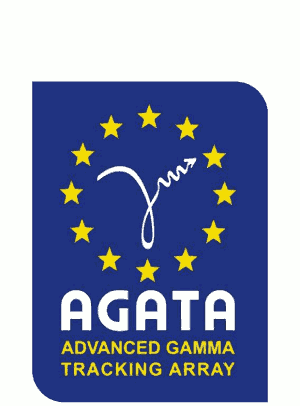}}
\maketitle
\begin{abstract}
\begin{onehalfspace}
The purpose of this master thesis is to simulate suitable nuclear
reactions for a commissioning experiment, to be performed with the
AGATA {\normalsize $\gamma$-ray} tracking spectrometer. The main
aim of the work is to find a reaction, which gives large Doppler effects
of the emitted $\gamma$ rays, with as small contribution as possible
due to the energy and angular spread of the nuclei emitting the $\gamma$
rays. Inverse kinematics heavy-ion (HI) fusion reactions of the type
(HI,$\gamma$), (HI, n) on proton and deuteron targets have been studied.
Target effects were investigated using the program TRIM in order to
determine the impact on the Doppler effects caused by energy and angular
straggling in the target material. The cross sections of a large number
of reactions of protons and deuterons on nuclei with mass numbers
in the range $A\approx20-100$ have been evaluated using the TALYS
reaction code. The fusion-evaporation reactions, $\mathrm{d\left(^{51}V,n\right)^{52}Cr}$
and $\mathrm{d\left(^{37}Cl,n\right)^{38}Ar}$ were simulated in detail
using the Monte Carlo code evapOR. The interactions in AGATA of the
$\gamma$ rays emitted in these reactions were simulated using \noun{Geant}4.
The energy resolution of the $\gamma$ rays after $\gamma$-ray tracking
and Doppler correction were determined as a function of the interaction
position resolution of the germanium detectors. The conclusion of
this work is that of the two reactions $\mathrm{d\left(^{51}V,n\right)^{52}Cr}$
is more suitable for an AGATA commissioning experiment.\end{onehalfspace}

\end{abstract}
\tableofcontents{}

\chapter{Introduction}

\begin{figure}[H]
\begin{centering}
\includegraphics[scale=0.6]{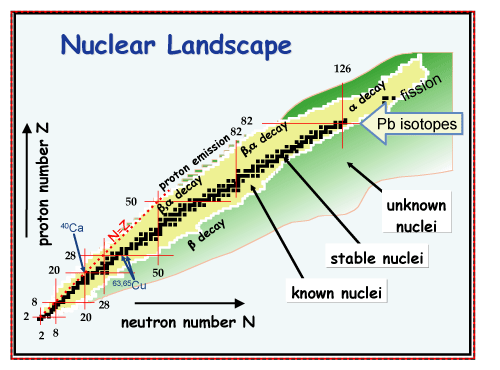}
\par\end{centering}

\caption{\label{fig:Chart}Chart of nuclides. (Figure from Holifield Radioactive
Ion Beam Facility, Oak Ridge \protect\url{http://www.phy.ornl.gov}) }

\end{figure}

Within the field of nuclear physics one attempts to describe how the
nucleus as a composition of nucleons can be held together, and how
the nucleons interact both individually and collectively inside the
nucleus. Nuclear models which describe the nuclear structure, can
be tested and further developed by improving nuclear detection methods.
The chart of nuclides, see figure \ref{fig:Chart}, currently consists
of about 3500 known nuclides. The green area of the figure \ref{fig:Chart}
shows the nuclides located within the drip-lines. According to theoretical
calculations less than half of the nuclides are known. In order to
produce and study the nuclear structure of these unknown nuclides,
great challenges face the development of precise detection instruments.
This master thesis aims to design a commissioning experiment to be
performed with the new $\gamma$-ray spectrometer, Advanced GAmma
Tracking Array (AGATA) \cite{2003RPPh...66.1095L}. The experiment
will be simulated and based on these calculations a possible reaction
for the upcoming experiments in 2009 will be proposed.

\section{Nuclear spectroscopy}

Spectroscopy early became an effective way of revealing information
about the atomic and subatomic quantal systems. In atomic physics,
studies of emission spectra corresponding to the transitions between
electronic orbits, was a key element in understanding the atom. Nuclear
physicists also use spectroscopic tools, by which one may retrieve
valuable information about the inner structure of the nuclei. The
energy levels (states) of the nuclei are denoted by their spin ($J$),
and parity ($\pi$), as $\left(J^{\pi}\right)$. The de-excitation
process of the states occurs either through emission of particles
$\left(\mathrm{e}^{-},\alpha\:,\mathrm{p}...\right)$ or $\gamma$
rays. The $\gamma$ radiation emitted by the de-exciting nucleus spans
an energy range from a few keV to tens of MeV. Gamma-ray spectroscopy
plays a pivotal role in discovering new exotic nuclei far beyond the
line of stability. Theoretical models can be evaluated and improved
by testing them with the experimental data obtained by using $\gamma$-ray
spectroscopic tools. Among the $\gamma$-ray spectroscopy instruments
used so far, the high-purity germanium detector (HPGe) is the most
successful one. By assembling a large number of such detectors in
an array surrounding the position of the nuclear reaction, one can
obtain efficient and precise detection of the emitted $\gamma$ rays. 

One major factor contributing to the uncertainty in the measured $\gamma$-ray
spectra is the Doppler effect. Considering in-beam experiments, the
nuclei emitting $\gamma$ radiation are usually not at rest in the
laboratory frame; hence the measured energy of the $\gamma$ rays
is subject to an uncertainty caused by the Doppler effect. If not
corrected for, this will lead to the broadening of the $\gamma$-ray
spectra. The Doppler effect depends on two factors, the velocity of
the nucleus and the angle between the direction of motion of the nucleus
and the emitted $\gamma$ ray. Due to the second factor, the first
interaction point of the $\gamma$ ray in the detector should be determined
in order to minimize the Doppler effects on the broadening of the
$\gamma$-ray peaks. One main advantage with new $\gamma$-ray spectrometers
compared with previous instruments, is the $\gamma$-ray tracking
technique, by which the first interaction point can be determined
with much better precision than before, which enhances the Doppler
correction capabilities. At present two such $\gamma$-ray spectrometers
are being developed, AGATA \cite{Agataproject} in Europe and GRETA
\cite{2003RPPh...66.1095L} in USA.

\section{Motivation }

It is of great interest to study how one can compensate for the Doppler
broadening of the $\gamma$-ray peaks in order to achieve a higher
energy resolution with the AGATA spectrometer. The task of this work
was to find a suitable nuclear reaction, in which the Doppler shifted
$\gamma$ rays can be studied using one AGATA triple cluster detector
in an experiment to be performed at INFN/LNL in Italy. 

\begin{figure}[H]
\begin{centering}
\includegraphics[scale=0.5]{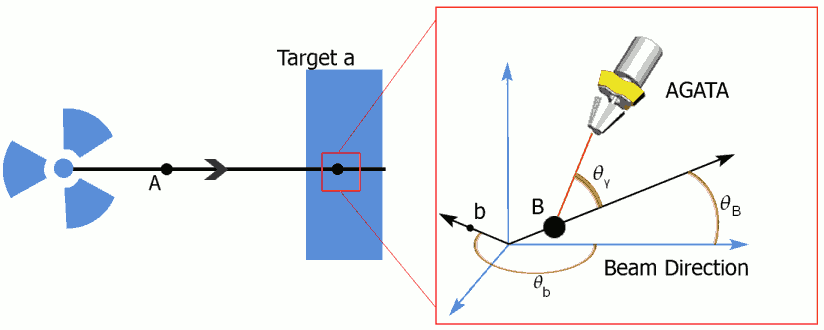}
\par\end{centering}

\caption{\label{fig:A-schematic-view}A schematic view of the experiment where
the ion $A$ from the accelerator collides with the target $a$. $\theta_{\mathrm{B}}$
and $\theta_{b}$ are the angle of emission of the residual nucleus
and the emitted particle respectively, and $\theta_{\mathrm{\gamma}}$
is the angle between of the direction of the residual nucleus B and
the emitted $\gamma$ ray. }

\end{figure}

One of the aims of this experiment is to use the Doppler effects to
determine the interaction position resolution of the AGATA germanium
detectors. Figure \ref{fig:A-schematic-view} shows a schematic view
of the experimental setup. The reaction in study is

\begin{equation}
A+a\rightarrow B+b,\end{equation}

where $A$ is the accelerated ion colliding with target $a$. The
products after the collision are the particle $b$ and the residual
nucleus $B$, which emits the Doppler shifted $\gamma$ ray.

\chapter{Theory}

\section{Nuclear models}

There are essentially two classes of nuclear models. The first class
is called \textit{independent particle models}, where the Pauli principle
restricts the collisions of the nucleons inside the nucleus, leading
to a large mean free path. The shell model belongs to this class.
The second class of models assumes strongly interacting nucleons inside
the nucleus, resulting in a small mean free path. The nuclear liquid
drop model belongs to this class. In general these models are called
\textit{collective models} as they attempt to describe phenomena that
involve the nucleus as a whole, e.g. vibrations and rotations. 

In this chapter brief descriptions of the shell model and of the collective
model are given.

\subsection{\label{sub:The-Shell-model}The shell model}

According to the Pauli principle the fermions cannot occupy the same
quantum state. Therefore nucleons can orbit almost freely inside the
nucleus, similarly to the electrons in the atom. In the atom, electrons
are moving in a central potential caused by the positively charged
nucleus. In the nucleus, the potential is created by the nucleons
themselves through the nucleon-nucleon interaction. In the shell model,
the nucleons move in a central potential usually parametrized as \begin{equation}
V\left(r\right)=\frac{V_{0}}{1+\mathrm{e}^{^{(r-R)/a}}}+V_{\mathrm{\ell s}}(r)\mathbf{\overrightarrow{\ell}}\cdot\mathbf{\mathrm{\overrightarrow{s}}}.\label{eq:wood}\end{equation}
The first term in equation \ref{eq:wood} is the \textit{Woods Saxon
potential} with the parameters: potential depth $V_{0}$, diffuseness
parameter $a$, nuclear radius $R=r_{0}A^{2/3}$ ($A$ is the nucleon
number and $r_{0}$ is the radius parameter). The second term is the
spin-orbit interaction, where the spin ($\overrightarrow{s}$) and
angular momentum ($\overrightarrow{\ell}$) vectors can be oriented
either parallel or anti-parallel in the total nucleon spin quantum
number $\overrightarrow{j}=\overrightarrow{\ell}\pm\frac{1}{2}$. 

The single-particle model is an extreme shell model form where the
nucleons move freely in a central potential without interacting with
each other. An odd-$A$ nucleus in this model is composed of an inert
even-even core, plus an unpaired nucleon. The last unpaired nucleon
determines the nuclear properties, such as spin and parity. All known
even-even nuclei have spin and parity $J^{\pi}=0^{+}$ in the ground
state. By introducing a residual force in the shell model, called
pairing, one can explain the ground state spin and parity of even-even
nuclei. The pairing interaction implies an attractive force between
pairs of nucleons with opposite orbital angular momenta in identical
orbits. The strongest attractive interaction occurs for nucleon pairs
with largest possible quantum numbers $m_{j}$ of opposite sign, i.e.
$m_{j}=+j$ and $m_{j}=-j$. The quantum number $m_{j}$ is the projection
of $j$ on a quantization axis and can have values from $-j$ to $j$.
The spin and parity of each pair of nucleons in identical orbits with
$\pm m_{j}$ is $0^{+}$. The attractive pairing interaction leads
to an increase of the binding energy for a completely paired nucleus
with total spin and parity $0^{+}$, compared with one with any number
of non-paired nucleons. 

The shell model can explain the presence of magic proton ($Z$) and
neutron ($N$) numbers in the nuclei. By including the spin-orbit
term in the potential, the shell model gives the magic numbers $Z,\: N=$
2, 8, 20, 28, 50, 82, 126, which corresponds to strongly bound nuclei
at filled shells. Experimental results verify these theoretical shell
model predictions. Doubly magic nuclei have magic proton and neutron
numbers. The few nuclei which are doubly magic are strongly bound,
like e.g. $^{4}\mathrm{He}$.

\subsection{The collective model}

Two major modes of collective excitations are described within this
model, namely vibrational and rotational excitations \cite{Jones}.

\subsubsection{Vibrational excitation of spherical nuclei }

The nucleus experiences oscillations which could either change the
size leaving the shape constant (\textit{breathing mode}), or change
the shape and leaving the density constant. The latter is more usual
and can be explained by different modes of oscillations. The surface
of a nucleus oscillating around a spherical equilibrium shape is written
as 

\begin{equation}
R(\theta,\phi,t)=R_{0}\left(1+\sum_{\lambda=0}^{\infty}\sum_{\mu=-\lambda}^{\lambda}\alpha_{\lambda\mu}(t)Y_{\lambda\mu}(\theta,\phi)\right),\end{equation}
where $\alpha_{\lambda\mu}(t)$ is the time dependent shape parameter,
$R_{0}$ is the radius of the sphere, and $Y_{\lambda\mu}(\theta,\phi)$
are the spherical harmonics. The multipolarity $\lambda$ determines
the oscillation type, with $\lambda=0$ corresponding to monopole
oscillations. These oscillations have no angular dependency. Dipole
oscillations occur when $\lambda=1$, causing the nucleus to vibrate
around a fixed laboratory reference point. The value of $\lambda=2$
implies quadrupole oscillations in which the nucleus alternates between
prolate and oblate shape. The energy levels of these vibrational modes
are given by

\begin{equation}
E_{N,\lambda}=\hbar\omega_{\lambda}\sum_{\mu=-\lambda}^{\lambda}\left(n_{\lambda\mu}+1/2\right)=\hbar\omega_{\lambda}\left(N_{\lambda}+\frac{1}{2}\left(2\lambda+1\right)\right),\end{equation}
where $N_{\lambda}$ is the number of phonons in the nuclear vibration.
A phonon is one quantum of vibrational energy, in analogy with the
photon being the electromagnetic quantum. The frequency is given by
$\omega_{\lambda}$. An increase of the number of phonons implies
higher energies, for a vibrational mode $\lambda$. These energies
will form a ladder with increasing $\hbar\omega_{\lambda}$. This
spectral phenomenon is called a \textit{vibrational band} \cite{hoistad}.

\subsubsection{Permanently deformed nuclei and rotational excitation }

The spherical symmetry is broken when moving away from closed shells.
Due to the pairing effect mentioned in \ref{sub:The-Shell-model},
the nucleons outside closed shells, will occupy the largest possible
$\pm m_{j}$ values, which leads to a non-spherical shape. The existence
of permanent deformations will open up the possibility for nuclear
rotations, which are not present in spherically symmetric nuclei.
Due to the rotation the nucleus feels a centrifugal stretching since
it is not a rigid body. In analogy with vibrational bands, rotations
form \textit{rotational bands. }

\section{\label{sec:Nuclear-reactions}Nuclear reactions}

Nuclear reactions can be \textit{elastic}, if the target and the projectile
are in the same states before and after the reaction (usually in their
ground states). When excited states are formed or new particles are
created, the collision is \textit{inelastic} \cite{Nuclearreactions}.
When nuclei collide with each other many reaction types can occur.
Direct reactions and compound reactions are two important types of
reactions.

\subsection{Direct reactions}

In a direct reaction the projectile can undergo stripping in which
case some of the nucleons are transferred from the projectile to the
target. An example of a stripping reaction is (d,p). A knockout reaction
can also occur, in which the incident nucleus knocks out nucleons
from the target nucleus, as in a (p,np) reaction. Direct reactions
occur on a very short time scale, $10^{-22}$ to $10^{-23}$s \cite{Jones},
and do not change the structure of the projectile and the target dramatically.

\subsection{\label{sub:Compound-reactions}Compound nucleus reactions}

According to \cite{Nuclearreactions}, compound nucleus reactions
occur when two nuclei form an intermediate state. The compound nucleus
formation between nuclei a and A is depicted as 

\begin{equation}
a+A\rightarrow X^{*}\rightarrow R^{*}+\sum\mathrm{particles,}\end{equation}
where $X^{*}$ is the excited compound nucleus and R is the excited
residual nucleus. The timescales of these reactions are relatively
long, roughly $10^{-16}$ to $10^{-18}$s. The information about the
constituents before the collision is permanently destroyed, hence
the output decay channels is independent of the compound nucleus formation.
If the same compound nucleus is reached through another reaction,
say $c+C\rightarrow X^{*}$with different colliding nuclei, the probability
of the decay channel $X^{*}\rightarrow R^{*}+\sum\mathrm{particles}$
remains the same. This hypothesis is called \textit{the independence
hypothesis} and it implies that it is possible to separate the cross
section $\sigma$, into the formation cross section $\sigma_{\mathrm{aA}}^{\mathrm{CN}}$
for projectile a and target A, and the branching ratio $G_{\mathrm{Bb}}^{\mathrm{CN}}$
(CN=compound nucleus) for the decay to particle b and nucleus B:

\begin{equation}
\sigma=\sigma_{\mathrm{Aa}}^{\mathrm{CN}}(E)G_{\mathrm{Bb}}^{\mathrm{C}\mathrm{N}}(E).\label{eq:cncrosssec}\end{equation}
The cross section term $\sigma_{\mathrm{Aa}}^{\mathrm{CN}}(E)$ is
the probability of the compound nucleus formation at the center of
mass energy $E$. As the compound nucleus must decay, the sum of all
the partial decays is $\underset{i}{\sum}G_{i}^{\mathrm{CN}}(E)=1$.
The excitation energy $E_{\mathrm{CN}}^{*}$ of the compound nucleus
can be expressed as

\nomenclature{equation exitation}{excitation}\begin{equation}
E_{\mathrm{CN}}^{*}=\frac{Z_{\mathrm{A}}+N_{\mathbf{\mathrm{A}}}}{Z_{\mathrm{A}}+N_{\mathrm{A}}+Z_{\mathrm{a}}+N_{\mathrm{a}}}E_{\mathrm{lab}}+\left[m_{\mathrm{a}}+m_{\mathrm{A}}-m_{\mathrm{CN}}\right]c^{2},\label{eq:exe}\end{equation}
where $E_{\mathrm{lab}}$ is the kinetic energy of the projectile
in the laboratory frame and the masses $m_{a}$, $m_{A}$ and $m_{CN}$
are the atomic masses of the projectile, target, and compound nucleus,
respectively. The compound nucleus de-excites through particle evaporation.

\subsubsection{\label{sub:Compound-Formation}Compound nucleus formation }

\begin{figure}[H]
\begin{centering}
\includegraphics{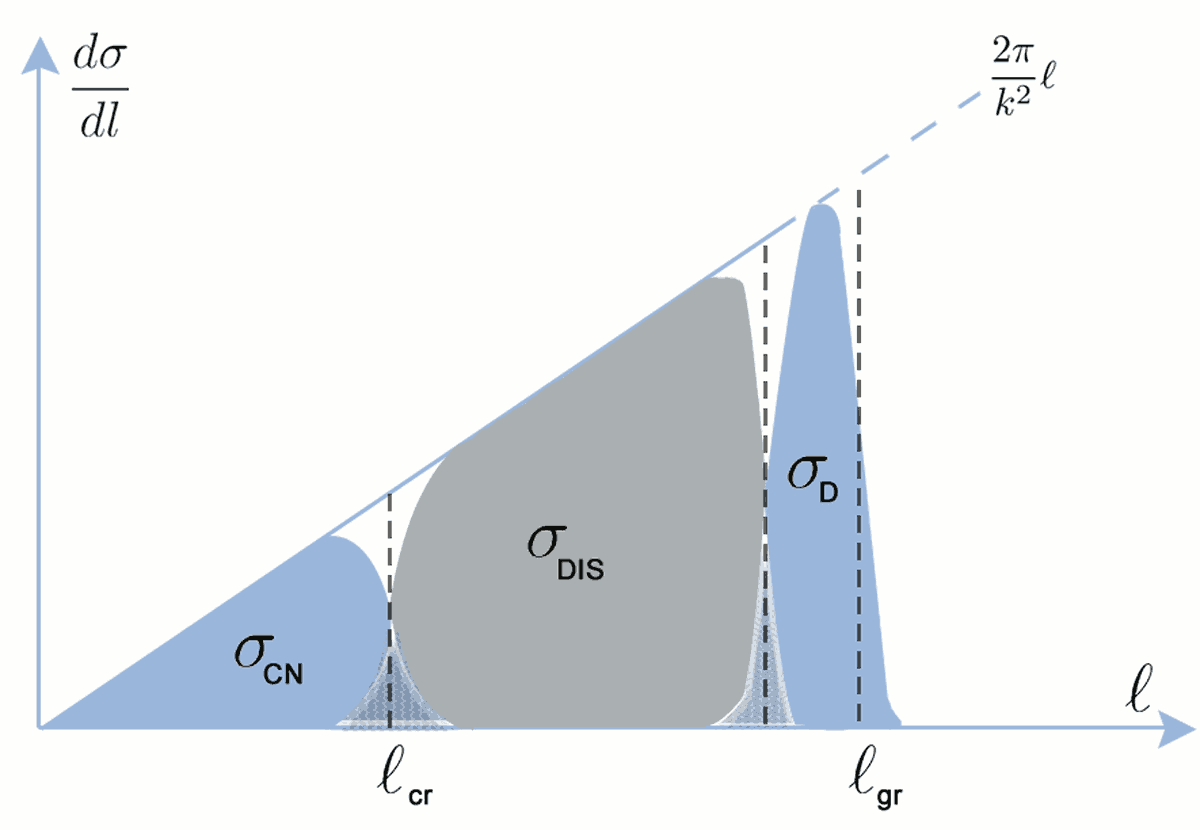}\caption{\label{fig:Differential-cross-section}Differential cross section
as a function of orbital angular momentum. }

\par\end{centering}

\end{figure}

Depending on the collision type and the impact parameter, high orbital
angular momenta may be transferred to populate the compound nucleus.
The maximum amount of angular momentum transfer in the compound nucleus
reaction, is denoted with $\ell_{\mathrm{cr}}$ \cite{Bock}. By considering
other reaction types, higher $\ell$ values can be reached. Figure
\ref{fig:Differential-cross-section} shows the differential cross
section $d\sigma/dl$ as a function of $\ell$. 

 Direct reactions occur for peripheral collisions implying high angular
momentum transfers. When the total angular momentum transfer $\ell_{\mathrm{gr}}$
is reached, the cross section drops. The \textit{sharp cut-off model
}states that this drops instantly when $\ell_{\mathrm{gr}}$ is reached.
This assumes a well-defined nuclear radius and a behavior similar
to the classical rigid-body collision. But since experiments show
that the nuclear radius is diffuse, and the border is not well-defined,
the collision may not fall drastically when $\ell_{\mathrm{gr}}$
is reached. Instead of this sharp cut-off model, a smooth drop-off
model is showed in figure \ref{fig:Differential-cross-section} \cite{Nuclearreactions,Bock}. 

The angular momentum transfer is essential in order to understand
the probabilities for nuclear reactions. According to the statistical
model, for heavy-ion collisions, the reaction cross section for the
compound formation depends on the transfer of orbital angular momentum
$\ell$ as \cite{Bock}

\begin{equation}
\sigma_{\mathrm{Aa}}^{\mathrm{CN}}\left(E\right)=\pi k^{-2}\sum_{0}^{\infty}\left(2\ell+1\right)T_{\ell}P_{\ell}^{\mathrm{CN}},\end{equation}
where $T_{\ell}$ is the transmission coefficient, $k$ is the wave
number, and $P_{\ell}^{\mathrm{CN}}$ is the probability of the $\ell$-wave
to enter the compound nucleus. In compound reactions the transmission
coefficient $T_{\ell}$ approximately equals unity for relatively
high angular momenta. The cross section can be simplified into 

\begin{equation}
\sigma_{\mathrm{Aa}}^{\mathrm{CN}}\left(E\right)=\pi k^{-2}\sum_{0}^{\ell_{\mathrm{cr}}}\left(2\ell+1\right)P_{\ell}^{\mathrm{CN}}.\end{equation}
The sum runs from orbital angular momentum 0 until the critical value
$\ell_{\mathrm{cr}}$ in which value the compound nucleus formation
cross section rapidly drops to zero. 

The cross section for final formation and decay is finally given by
\cite{statmodel}  

\begin{equation}
\sigma_{Aa\rightarrow i}=\sum_{J_{\mathrm{c}}}\sigma_{\mathrm{Aa}}^{\mathrm{CN}}(E)\frac{\Gamma_{i}(E_{i},J_{\mathrm{c}})}{\sum\Gamma_{j}(E_{j},J_{\mathrm{c}})}=\sum_{J_{\mathrm{c}}}\left(\pi k^{-2}\sum_{S=|I-s|}^{I+s}\sum_{\ell=|J_{\mathrm{c}}-S|}^{J_{\mathrm{c}}+S}\frac{\left(2J_{c}+1\right)}{(2s+1)(2I+1)}T_{\ell}(E)\right)\frac{\Gamma_{i}}{\sum\Gamma_{j}},\end{equation}
 where $\sigma_{\mathrm{aA}}\left(E\right)$ is the compound nucleus
formation cross section, dependent on the entrance channel-energy
$E$. The decay width $\Gamma_{i}$ is for a specific final state
$i$ in the exit channel. The target spin is given by $I$, and S
is the channel spin. The projectile spin is represented by $s$.

\subsubsection{Compound nucleus decay}

\begin{figure}[H]
\centering{}\includegraphics{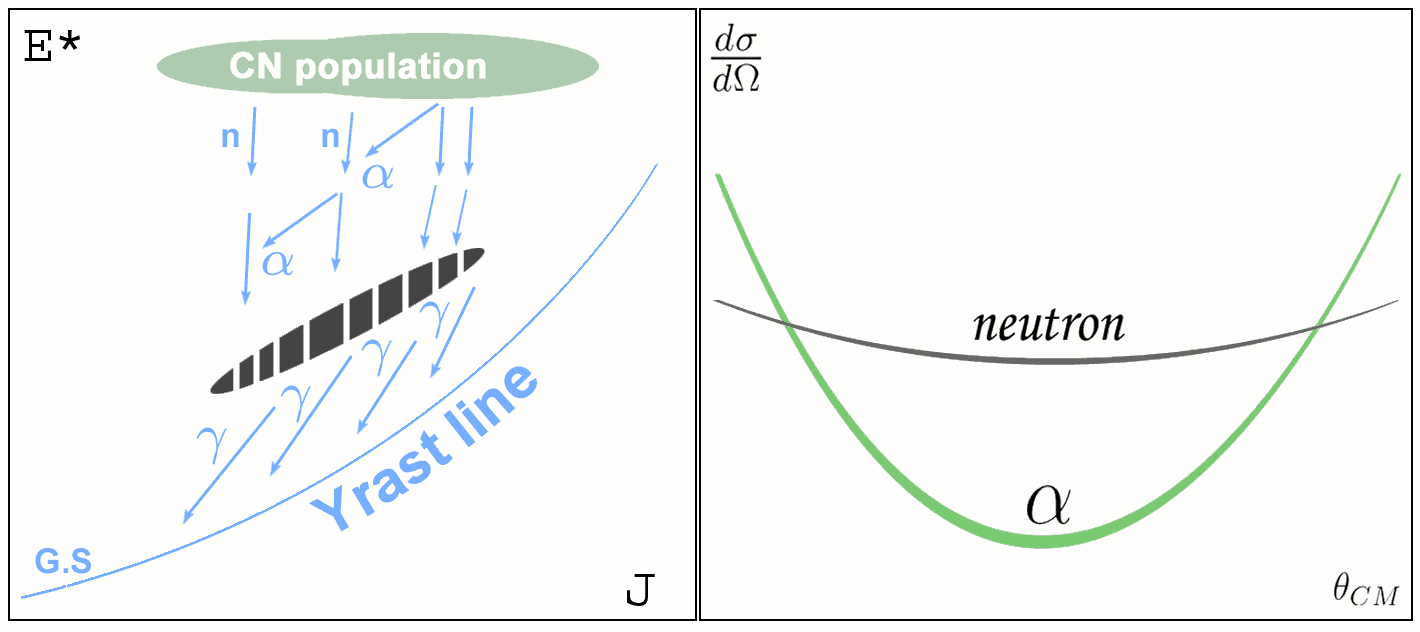}\caption{\label{fig:In-this-population}In the left figure a population scheme
is shown, in which compound nuclei are created with a given excitation
energy $E^{*}$. Particle evaporation releases most of the energies,
followed by $\gamma$ ray cascades to finally reach the ground state.
In the right figure the angular distribution of neutrons and $\alpha$
particles are schematically presented, indicating an increased anisotropy
for the $\alpha$ particles. }

\end{figure}

When the compound nucleus is formed it populates energy levels high
above the ground state. The nucleus then tries to reach the ground
state through an evaporation process in which mainly light particles
(neutrons, protons, $\alpha$ particles) are subsequently emitted
in a so called cascade, see figure \ref{fig:In-this-population}.
While charged particles, such as protons and $\alpha$ particles,
have to overcome the Coulomb barrier in order to escape the excited
nucleus, neutral particles do not. This will be reflected in the energy
distribution of the emitted particles: neutrons are emitted with lower
kinetic energies. See figures \ref{fig:angular effect doppler} and
\ref{fig:a.-alpha} for a comparison of energy distributions of neutrons
and alpha particles. 

Alpha particles carry away larger values of angular momentum from
the compound nucleus than neutrons and protons, as shown in figure
\ref{fig:In-this-population}. This follows since 

\begin{equation}
\mathbf{\overrightarrow{\ell}}=\overrightarrow{\mathrm{\mathit{r}}}\times\overrightarrow{\mathrm{\mathit{p}}}=\overrightarrow{\mathrm{\mathit{r}}}\times\mathit{\mathrm{\mathit{m}}\overrightarrow{\mathrm{\mathit{v}}}}\end{equation}
 and both the mass $m$, and the velocity $v$, is higher for $\alpha$
particles than for neutrons and protons. \\

\label{sub:Energy-distribution-of}\textbf{Energy distribution of
evaporated particles} \\
The energy distributions of the evaporated particles have a Maxwell-Boltzmann
shape, according to the spin dependent statistical model. Following
the derivation of Lang and Lecouteur \cite{Lang}, the probability
of emission is 

\begin{equation}
W_{\mathrm{p}}\left(E^{*},J,E_{\mathrm{R}}^{*},J_{\mathrm{R}}\right)=\frac{2S_{\mathrm{p}}+1}{\pi^{2}\hbar^{3}}\mu E_{\mathrm{p}}\sigma_{\mathrm{inv}}^{J}(E^{*})\frac{\rho(E_{\mathrm{R}}^{*},J_{\mathrm{R}})}{\rho(E^{*},J)},\label{eq:WW}\end{equation}
where the emitted particle has a reduced mass $\mu$, energy $E_{\mathrm{p}}$,
and spin $S_{\mathrm{p}}$. The inverse compound nucleus production
cross section is $\sigma_{\mathrm{inv}}^{J}(E^{*})$, and $\rho(E^{*},J)$
and $\rho(E_{\mathrm{R}}^{*},J_{\mathrm{R}})$ are the level densities
of the initial and residual nuclei at spin and excitation energy $E^{*}$,
$J$ and $E_{R}^{*}$, $J_{R}$, respectively.\\

\textbf{Angular distribution of evaporated particles}\\
Based on the work of Ericson and Strutinsky \cite{ericson}, the
orbital angular momentum of the incident particle is orthogonal to
the direction of the beam. The total angular momentum aligns preferably
in the plane perpendicular to the beam axis. The angular distribution
is symmetric with respect to the angle $90^{\circ}$ relative to the
direction of the incoming beam. According to \cite{Bock,ericson},
the angular distribution of the evaporated particles can be written
as 

\begin{equation}
W_{J,\ell}(\theta_{\mathrm{p}},E_{\mathrm{p}})=\frac{1}{4\pi}\sum_{k}^{\infty}(-1)^{k}(4k+1)\left(\frac{(2k)!}{(2^{k}k!)}\right)^{2}j_{2k}\left(\frac{iJ\ell}{\sigma_{\mathrm{f}}^{2}}\right)P_{2k}(\cos\theta_{\mathrm{p}}).\label{eq:angulardist}\end{equation}
The spherical Bessel functions $j_{2k}$ are of order $2k$ and $\theta_{\mathrm{p}}$
is the angle between the incident beam and the direction of the emitted
particle in the center of mass system. The spin cut-off parameter
of the residual nucleus is given by $\sigma_{\mathrm{f}}$. The Legendre
polynomial $P_{2k}$ is even, thus there is a symmetry around $90^{\mathrm{o}}$.
Figure \ref{fig:In-this-population} illustrates that $\alpha$ particles
are emitted mostly at large ($\sim180^{\circ}$) and small ($\sim0^{\circ}$)
angles. The reason for this is that they remove large values of the
angular momentum $\ell$ when evaporated from the compound nucleus.
For neutrons the evaporation is nearly isotropic. See figure \ref{fig:sintheta}
for an angular distribution of neutrons obtained by the simulations
performed in this work.

\chapter{\label{cha:Experimental-Nuclear-Physics}Nuclear spectroscopy methods}

\section{Interaction of $\gamma$ rays with matter}

When $\gamma$ rays enter matter, three interaction types can occur,
photoelectric effect, Compton scattering and pair production.

\subsection{\label{sub:Photoelectric-Effect}Photoelectric effect}

When irradiating a metal surface with light above a certain frequency,
an electric current will flow. Einstein found that electrons are removed
from the metal and that the photon is absorbed by the atom \cite{photoelectric}.
The photo-electron can only be ejected when the photon energy is higher
than the binding energy of the electron. The kinetic energy for the
photoelectron is given by 

\begin{equation}
E_{\mathrm{e}}^{\mathrm{kin}}=E_{\gamma}-B_{\mathrm{e}},\end{equation}
where $E_{\gamma}$ is the photon energy and $B_{e}$ is the electronic
binding energy. The photoelectric effect dominates at low $E_{\gamma}$
(see figure \ref{fig:Domains-of-electromagnetic}).

\subsection{\label{sub:Compton-Scattering}Compton scattering}

When the photon energy increases, the probability of photon scattering
with electrons increases (see figure \ref{fig:Domains-of-electromagnetic}).
According to Compton in his article from 1923 \cite{Compton}, the
observed photon wavelength $\lambda$ increases after the interaction,
leading to a decreasing energy, since $E=hc/\lambda$. Recoil energy
is transferred from the photon to the electron. Due to energy and
momentum conservation, one arrives at the Compton scattering formula
which gives the energy of the scattered $\gamma$ ray

\begin{equation}
E'_{\gamma}=\frac{E_{\gamma}}{1+(E_{\gamma}/mc^{2})(1-\cos\theta_{\mathrm{C}})}.\label{eq:compton}\end{equation}
Here $E_{\gamma}$ is the initial energy of the $\gamma$ ray, and
$m$ is the mass of the electron. the scattering angle of the photon
is $\theta_{\mathrm{C}}$.

\subsection{\label{sub:Pair-Production}Pair production}

If the photon energy is large enough, pair production may occur, i.e.
creation of an electron-positron pair, as $\gamma\rightarrow\mathrm{e}^{+}+\mathrm{e}^{-}$.
The minimum energy required is 1.022 MeV. Due to momentum and energy
conservation this process is only possible if it occurs nearby a particle
with non-zero rest mass, e.g. a nucleus \cite{Krane}. \\

\begin{center}
\begin{figure}[H]
\begin{centering}
\includegraphics{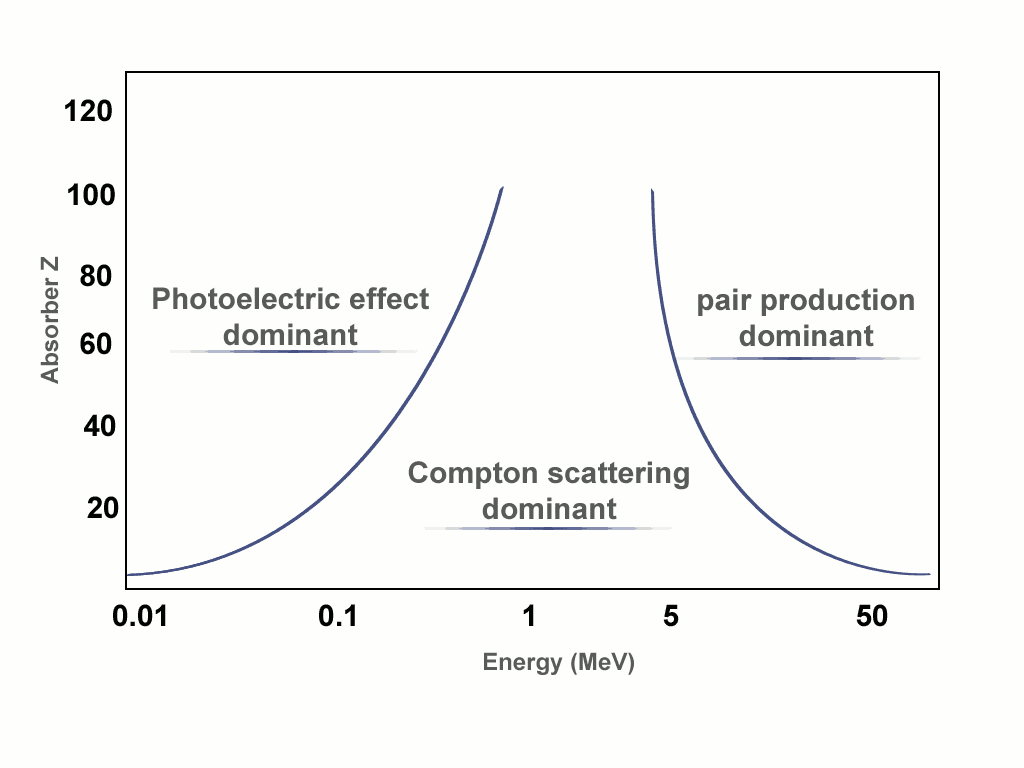}
\par\end{centering}

\caption{\label{fig:Domains-of-electromagnetic}Domains of electromagnetic
interaction with matter. \cite{Krane}}

\end{figure}

\par\end{center}

\subsection{Attenuation coefficients}

The intensity of a beam passing through matter, decreases exponentially 

\begin{equation}
I=I_{0}e^{-\mu\delta},\end{equation}
where $I_{0}$ is the incident beam intensity, $\mu$ is the total
attenuation coefficient and $\delta$ is the material thickness. The
probability for a photon removal per unit length is given by $\mu$.
The total attenuation coefficient is the sum of the attenuation coefficients
for the photoelectric effect $\left(\tau\right),$ Compton scattering
$\left(c\right)$ and pair production $\left(\kappa\right)$

\begin{equation}
\mu=\tau+c+\kappa.\end{equation}

The inverse of the attenuation coefficient is equal to the mean free
path of the $\gamma$ rays 

\begin{equation}
\lambda_{\mathrm{m}}=\frac{1}{\mu}.\end{equation}

\section{Interaction of ions with matter}

Transport of ions in matter is mainly affected by electromagnetic
interactions with the electrons of the material. Rutherford scattering,
i.e. scattering between nuclei, is rare in comparison with the interaction
with the electrons. The energy loss of the ion per distance traveled
is given by \cite{Bethebloch} 

\begin{equation}
-\frac{\mathrm{d}E}{\mathrm{d}x}=\left(\frac{\mathrm{e}^{2}}{4\pi\epsilon_{0}}\right)^{2}\frac{4\pi z^{2}N_{0}Z\rho}{mv^{2}A}\left[\ln\left(\frac{2mv^{2}}{I\left(1-\left(v/c\right)^{2}\right)}\right)-\frac{v^{2}}{c^{2}}\right],\label{eq:electrons}\end{equation}
where $A,$ $Z$, $z,$ $E$ and $v$ are the mass number, atomic
number, electric charge, energy and velocity of the ion respectively,
$m$ is the electron mass, $\rho$ the density of the target, $I$
is the mean excitation potential of the target, and \textbf{\textsc{$N_{0}$}}Avogadro's
number.

\section{Doppler Effects}

\begin{figure}[H]
\begin{centering}
\includegraphics[scale=0.25]{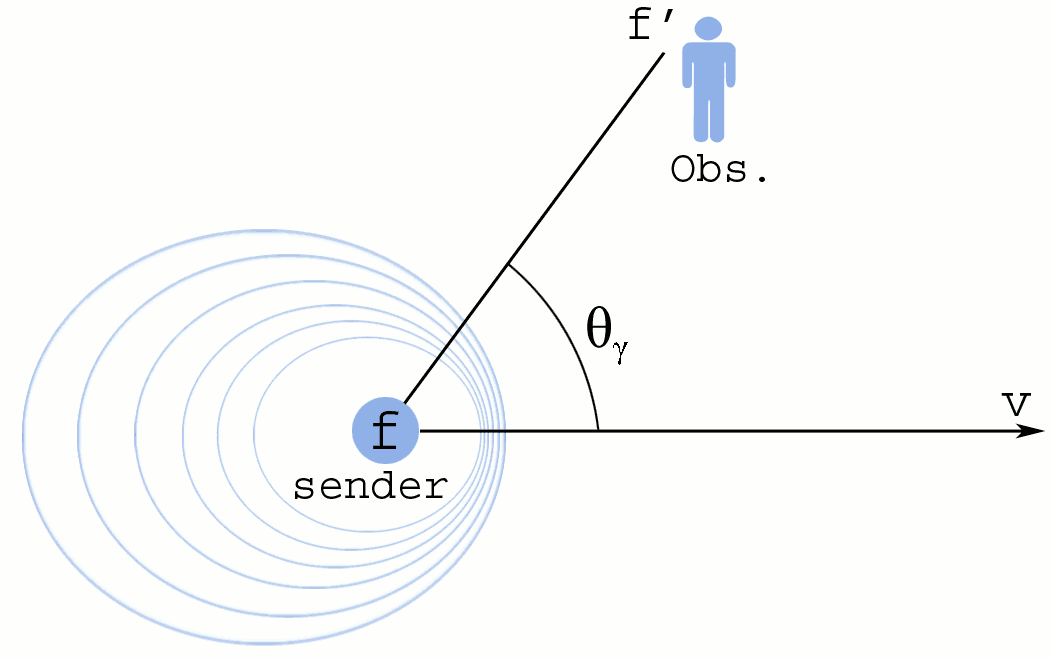}
\par\end{centering}

\caption{\label{fig:Doppler-shift}Illustration of the Doppler shift. }

\end{figure}

Doppler Effects are an important concept for nuclear spectroscopy.
Gamma radiation emitted by a nucleus in motion, appears to have a
different energy than the actual transition energy. The frequency
of the $\gamma$ rays is lower if the nucleus moves away from the
observer, and higher when it moves towards the observer. Figure \ref{fig:Doppler-shift}
demonstrates the change of frequency $f,$ which is observed by the
receiver as $f'$.

The relativistic version of the Doppler shift formula for a source
moving with velocity $v$ is \cite{Feynman} 

\begin{equation}
E_{\gamma}=E_{\gamma_{0}}\frac{\sqrt{1-\frac{v^{2}}{c^{2}}}}{1-\frac{v}{c}\cos\theta_{\gamma}},\label{eq:doppler2}\end{equation}
where $E_{\gamma}$ and $E_{\gamma_{0}}$ $(E=hf)$ are the measured
$\gamma$-ray energy and the $\gamma$-ray transition energy, respectively,
and $\theta$ is the angle between the receiver and the direction
of motion of the source. The velocities in the simulations of the
AGATA commissioning experiments are below 10 \% of the speed of light,
hence the relativistic factor $\sqrt{1-v^{2}/c^{2}}$ is very close
to unity. In this case a simplified non-relativistic approximation
of equation \ref{eq:doppler} is appropriate to use:

\begin{equation}
E_{\gamma}=E_{\gamma_{0}}\left(1+\frac{v}{c}\cos\theta_{\gamma}\right).\label{eq:doppler}\end{equation}

\section{\label{sec:Detector-physics}Gamma-ray spectroscopy}

\begin{figure}[H]
\begin{centering}
\includegraphics[scale=0.5]{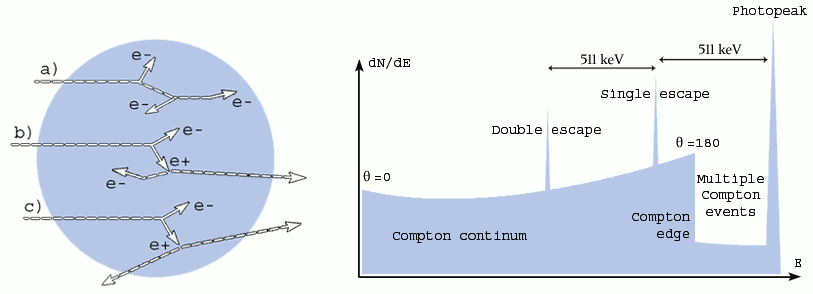}
\par\end{centering}

\caption{\label{fig:Three}Three $\gamma$-rays enter the detector in the left
figure. a) Two Compton scatterings followed by a photoelectric effect.
b) A pair production followed by single escape. c) Pair production
followed by double escape. The figure on the right shows the main
characteristics of a $\gamma$-ray spectrum. }

\end{figure}

When photons enter a $\gamma$-ray detector they may deposit all or
part of their energy in the detector. Figure \ref{fig:Three} shows
how different interactions contribute to a typical $\gamma$-ray spectrum.

Photons depositing all their energy in the detector give rise to the
full-energy peak (photopeak) while the ones that do not deposit all
their energy, but escape after one or more Compton scatterings, contribute
to the Compton continuum. The scattering angle in equation \ref{eq:compton},
determines the energy deposited. Back-scattering of the photons forms
the Compton edge in figure \ref{fig:Three}. The amount of $\gamma$
rays that escape from the detector depends on the size of the detector
compared to the mean free path of the photons in the detector.

\subsection{HPGe detector and energy resolution }

The high-purity germanium (HPGe) detector is undoubtedly the most
successful instrument for spectroscopic studies of the nuclear structure.
The superior performance of the HPGe detector is mainly due to its
excellent energy resolution, which allows the distinction of $\gamma$
rays with close lying energies. The energy resolution is characterized
by the full width at half maximum (FWHM) of the $\gamma$-ray peaks.
Two factors contribute to the FWHM, here labeled $W$ \cite{knoll}: 

\begin{equation}
W=\sqrt{W{}_{\mathrm{i}}^{2}+W_{\mathrm{D}}^{2}}.\label{eq:FWHM}\end{equation}
The intrinsic width $W_{\mathrm{i}}$ is caused by the properties
of the detector itself and consists of three terms: the electronic
noise $W_{\mathrm{en}}$, which is independent of the measured $\gamma$-ray
energy, the statistical factor $W_{\mathrm{sf}}$, which depends on
the square root of the energy, and the incomplete charge collection
$W_{\mathrm{cc}}$, which is proportional to the energy, 

\begin{equation}
W_{\mathrm{i}}=\sqrt{W{}_{\mathrm{en}}^{2}+W{}_{\mathrm{sf}}^{2}+W{}_{\mathrm{cc}}^{2}}.\label{eq:310}\end{equation}
The second factor $W_{D}$ is due to the Doppler effects and it is
non-zero only if the $\gamma$ ray is emitted by a moving source.
The Doppler factor consists of three terms:

\begin{equation}
W_{\mathrm{D}}=\sqrt{W{}_{\theta_{\mathrm{\gamma}}}^{2}+W{}_{\mathrm{\theta_{r}}}^{2}+W{}_{v_{\mathrm{r}}}^{2}}.\label{eq:311}\end{equation}
$W{}_{\theta_{\gamma}}$ is due to the broadening caused by the uncertainty
in the determination of the emission angle of the $\gamma$ ray, $\theta_{\gamma}$.
This term depends on the position resolution of the first interaction
point in the detector. The terms $W{}_{\mathrm{\theta_{r}}}$ and
$W{}_{v_{\mathrm{r}}}^{}$ are due to the uncertainty in the determination
of the angle and velocity, respectively, of the residual nucleus.

\section{AGATA}

AGATA is a $\gamma$-ray spectrometer which in its final $4\pi$ configuration
will consist of 180 HPGe crystals assembled in 60 triple cluster detectors
(see figure \ref{fig:High-purity-Germanium}). Each HPGe crystal is
of type closed-end coaxial. The crystals are 9 cm long, have circular
shape with a diameter of 8 cm at the rear end, and a hexagonal shape
in the front. The crystals are also electrically segmented into 6
longitudinal and 6 azimuthal segments which together with the core
gives 37 separate signals per crystal. The crystals are encapsulated
in aluminum cans (see figure \ref{fig:High-purity-Germanium}). 

\begin{center}
\begin{figure}[H]
\begin{centering}
\includegraphics{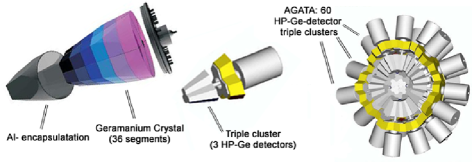}
\par\end{centering}

\caption{\label{fig:High-purity-Germanium}The AGATA HPGe detectors \cite{Agataproject}. }

\end{figure}

\par\end{center}

AGATA will use advanced digital electronics and pulse-shape analysis
(PSA) techniques to determine the 3D position and the energy deposition
in each $\gamma$-ray interaction point. An example of the PSA technique
is shown in figure \ref{fig:The-pulse-shape}. The interaction points
determined by the PSA are then fed into a $\gamma$-ray tracking algorithm,
which is based on the Compton scattering formula, equation \ref{eq:compton},
and by which the full-energy and the first interaction point of each
$\gamma$ ray is reconstructed. 

\begin{center}
\begin{figure}[H]
\begin{centering}
\includegraphics[width=14cm]{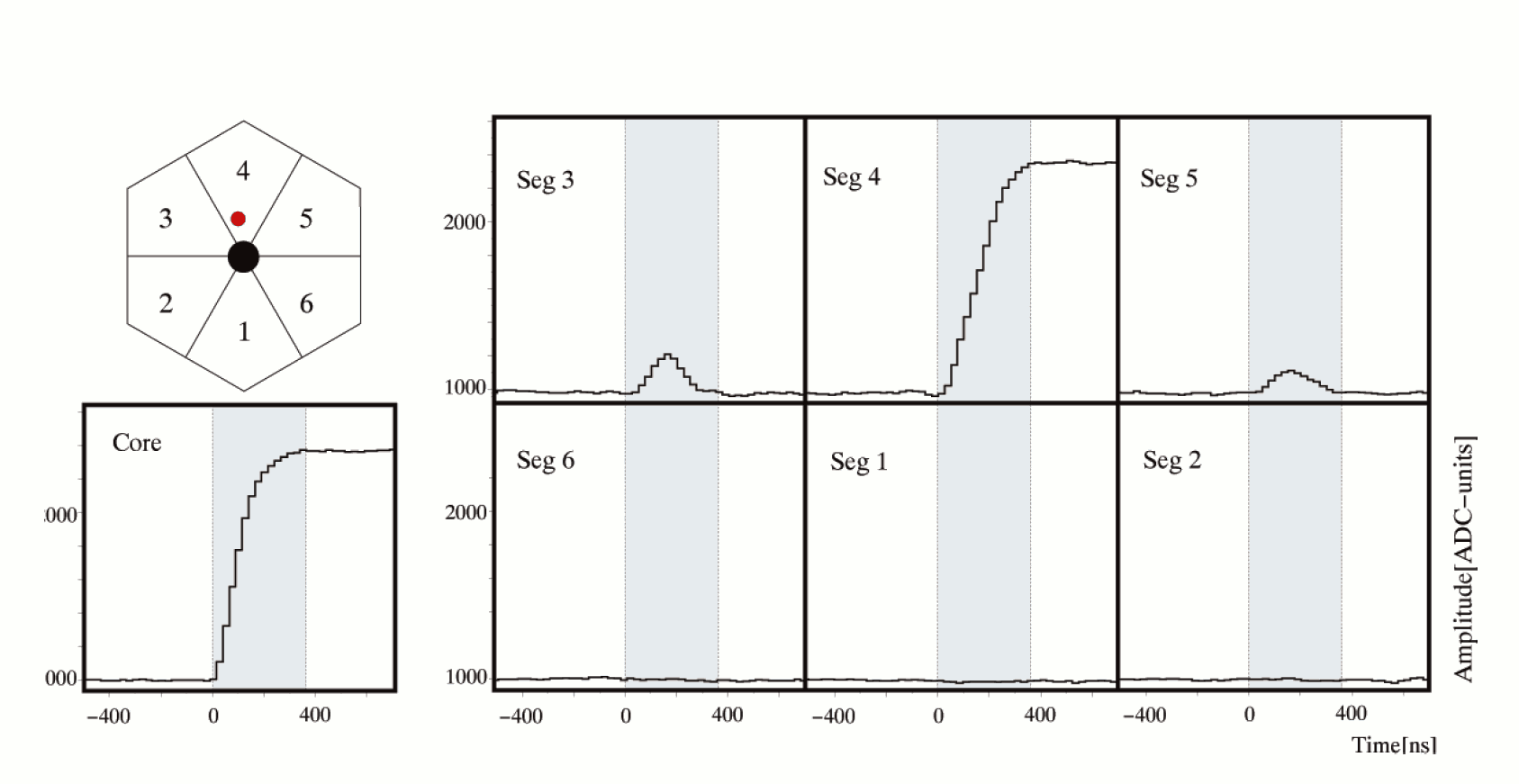}
\par\end{centering}

\centering{}\caption{\label{fig:The-pulse-shape}Illustration of the pulse shape technique.
In this event a hit in segment 4 can be seen. A signal is registered
from the core and two mirror charges are seen in segments 3 and 5
\cite{Lopez-Martens_NIM_A533}.}

\end{figure}

\par\end{center}

The improvement to be expected by AGATA compared to earlier Ge-detector
arrays, not based on $\gamma$-ray tracking, will result in a larger
full-energy efficiency and much better Doppler correction capabilities.
With previous detector arrays one could only determine which detector
was hit by a $\gamma$ ray. Thus, one had an opening angle corresponding
to the diameter of the detector, typically $\sim5$ cm. In AGATA the
position resolution will be about 5 mm which will lead to a much smaller
value of $W_{\theta_{\gamma}}$ in equation \ref{eq:311} and, thus,
too much narrower $\gamma$-ray peaks. 

\begin{figure}[H]
\centering{}\includegraphics[width=9cm]{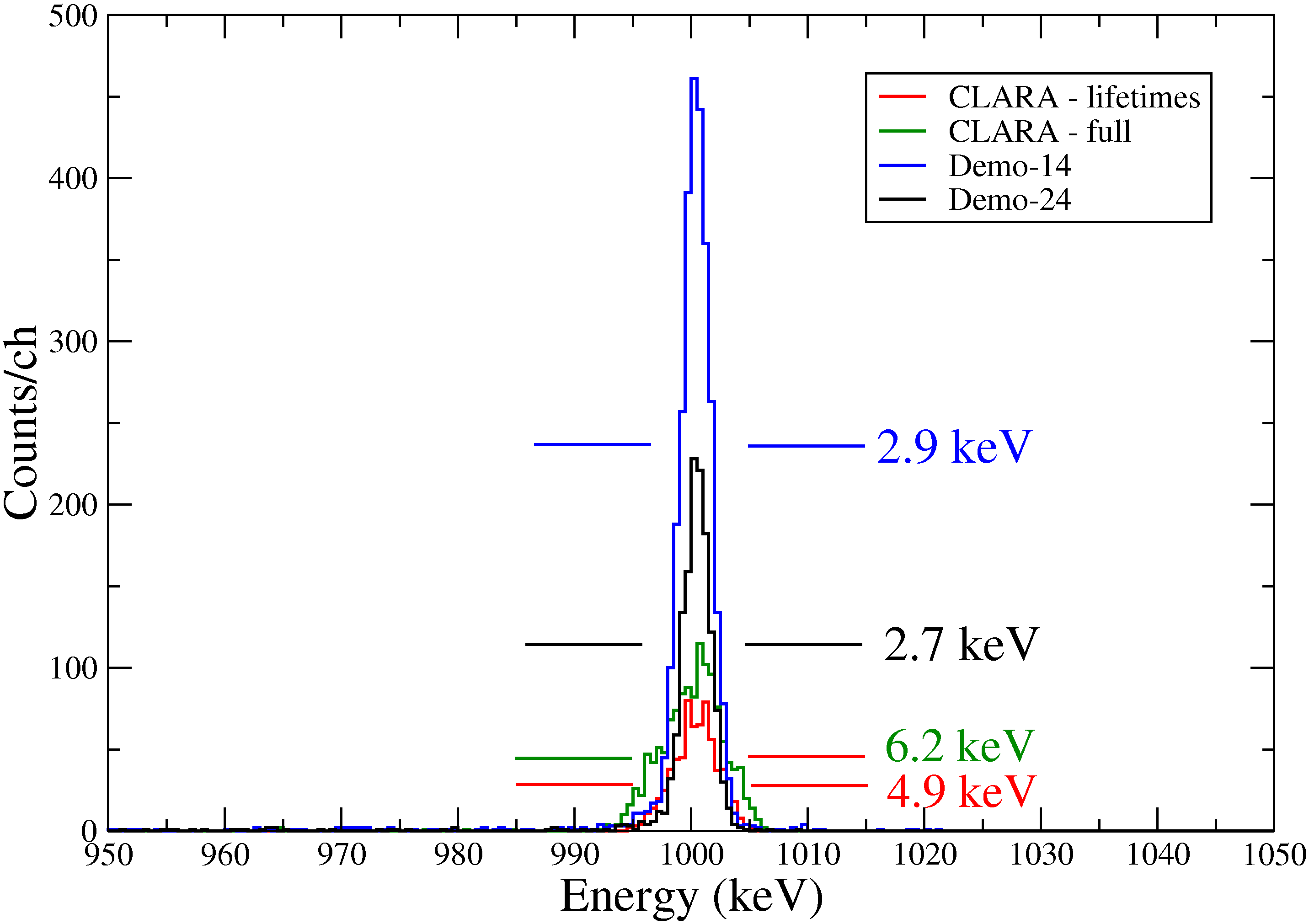}\caption{\label{fig:The-predicted-performance}The predicted performance of
the AGATA demonstrator (15 HPGe crystals) compared to the CLARA spectrometer,
which consists of about 100 non-segmented HPGe crystals. The FWHM
at a $\gamma$-ray energy of 1 MeV is more than a factor of two better
with AGATA (2.7 keV and 2.9 keV) at a target to detector front face
distance of 24 and 14 cm, respectively, than with the full CLARA array
(6.2 keV) \cite{farnea}. }

\end{figure}

AGATA will serve as a powerful tool to understand nuclear structure
far from the $\beta$-stability line. Several detection difficulties
such as Doppler broadening will be possible to compensate for using
the AGATA spectrometer. By having a better resolution of the full-energy
peaks, weak transitions in exotic nuclei can be studied \cite{Agataproject}.
Figure \ref{fig:The-predicted-performance} displays the predicted
performance of the AGATA demonstrator compared to the CLARA spectrometer.

\chapter{\label{sub:Requirements}Requirements}

The selection of reactions to be studied in this work was based on
the following list of criteria.
\begin{itemize}
\item Large Doppler shifts: This is required in order to study the interaction
position resolution of the AGATA detectors. The velocity of the residual
nucleus should be higher than 5 \% of the speed of light. 
\item Large reaction rate: The reaction rate is given by \begin{equation}
R=\frac{I}{q}\frac{\rho t\sigma}{m_{A}N_{\mathrm{A}}},\end{equation}
where $I$ is the ion beam current, $q$ the electric charge of the
ions, $t$ and $\rho$ the target thickness. and density, respectively,
$\sigma$ the cross section, $m_{A}$ the atomic weight, and $N_{\mathrm{A}}$
Avogadro's number. 
\item Available ion beams at LNL: The available beams at the LNL accelerator
laboratory are listed in appendix \ref{cha:Available-ion-beams}.
\item Competing decay channels: Different particle evaporation channels
compete. Proton and $\alpha$ particle evaporation give large recoil
energies to the residual nucleus, since they need high kinetic energy
to penetrate the Coulomb barrier. Neutrons do not face the same situation,
hence the average recoil energy given to the nucleus is smaller.
\item Position of $\gamma$-ray emission: In order not to have $\gamma$-ray
emission inside the target, the life times must be longer than the
time it takes for the ions to pass the target material. To have a
negligible influence on the Doppler effects, the life times must be
short enough so that the decay occurs within 1 mm after the target.
\item Energy and angular straggling of the ions in the target should be
as small as possible. 
\item Gamma-ray energies: A minimum $\gamma$-ray energy of 500 keV is required.
Lower energies will give smaller Doppler shifts, which are harder
to study. 
\item Strong $\gamma$-ray transitions: In even-even nuclei a strong $2^{+}\rightarrow0^{+}$
transition is often observed, therefore even-even final nuclei are
preferred. 
\end{itemize}
In this work two reaction types were studied in order to find a suitable
reaction: a) proton capture $\mathrm{p}\left(X,\gamma\right)$ or
deuteron capture $\mathrm{d}\left(X,\gamma\right)$ and b) fusion
evaporation with emission of one neutron followed by $\gamma$ rays,
for example $\mathrm{p}\left(X,\mathrm{n}\gamma\right)$ or $\mathrm{d}\left(X,\mathrm{n}\gamma\right)$.
Both type of reactions were of inverse kinematics type, with a heavy
ion beam on a proton or deuteron target.

\chapter{Simulations}

\section{TALYS}

The code used to evaluate the cross sections was TALYS, a comprehensive
nuclear reaction modeling code \cite{nd2007_talys}. The projectiles
and ejectiles in TALYS can be of type $\gamma$ rays, neutrons, $^{1-3}\mathrm{H}$,
$^{3-4}\mathrm{He}$, with energies in the range 0.1 to 200 MeV. The
mass number of the target can be in the range $A$=12 to $A$=339.
Note that by TALYS it is not possible to simulate the reactions in
inverse kinematics. TALYS was, however, used for calculating the cross
sections of the reactions, which are independent of the reaction being
of normal or inverse kinematics type. In this work the results obtained
with TALYS were in several cases compared to experimental data taken
from the National Nuclear Data Center. 

The total probability of a reaction to happen, i.e. the total cross
section $\sigma_{\mathrm{tot}}$, is the sum of the elastic $\sigma_{\mathrm{e}}$
and the inelastic cross section $\sigma_{\mathrm{ie}}$:

\begin{equation}
\sigma_{\mathrm{tot}}=\sigma_{\mathrm{e}}+\sigma_{\mathrm{ie}}.\end{equation}
The inelastic cross section is a sum of the compound nucleus, pre-equilibrium
and direct cross sections. The compound nucleus formation cross section
$\sigma_{\mathrm{cf}}$, is defined as the difference between the
reaction $\sigma_{\mathrm{reac}}$ and the direct cross section $\sigma_{\mathrm{direct}}$:

\begin{equation}
\sigma_{\mathrm{cf}}=\sigma_{\mathrm{reac}}-\sigma_{\mathrm{direct}}.\end{equation}
The output of TALYS contains the calculated cross sections for non-elastic
direct reactions, non-elastic and elastic compound nucleus reactions,
and non-elastic pre-equilibrium reactions.

\section{Target effects}

The target that will be used in the experiment has an effect on the
velocity and direction of the ions. The program used for studies of
the target effects was TRIM \cite{srim}, which is a Monte Carlo based
program that simulates collisions between incident ions and atoms
in the target material. Angular and energy distributions of the incident
ions are obtained from the TRIM simulations. The target material can
be chosen from a list of chemical substances and compounds.

\subsection{\label{sub:Sources-of-target}Sources of target effects}

When the ions enter the target material they are deflected due to
the electromagnetic interaction primarily with the electrons. Thus,
the ions will loose energy and deviate from the beam direction, which
leads to so called energy and angular straggling of the ions. There
are two main aspects of the target effects in the simulations. 
\begin{enumerate}
\item \textbf{Position of the reaction in the target}. If the reaction occurs
in the beginning of the target, the residual nucleus will experience
energy and angular straggling, whereas if the reaction occurs at the
end, the incident ion instead of the residual nucleus will lose kinetic
energy. The difference both in kinetic energy and atomic number of
the incident ion and residual nucleus leads to the difference in straggling. 
\item \textbf{Neutron energy and angular distribution}. After the reaction
the excited compound nucleus is formed, and particles are emitted.
Depending on the energy and angular distribution of the particles,
the residual nucleus will obtain different kinetic energies. The recoil
due to the neutron evaporation will either boost the nucleus or slow
it down. Since this happens inside the target material the energy
and angular straggling will change after the reaction. Two extreme
cases were investigated, one where the neutron is emitted at about
$0^{\mathrm{o}}$ relative to the ion beam, and the other where it
is emitted in at about $180^{\mathrm{o}}$. 
\end{enumerate}

\section{Fusion-evaporation simulation}

The fusion-evaporation simulations were done using the Monte Carlo
code evapOR \cite{beene}. This program allows for fusion of two arbitrary
nuclei, followed by evaporation of particles which can be of 7 different
types $(\mathrm{n,p,d,a,t,^{3}He,^{6}Li})$ as well as emission of
$\gamma$ rays. It uses the statistical model for compound nucleus
reactions, described in section \ref{sub:Compound-reactions}. 

In evapOR the velocity of the emitted particle in the laboratory system,
$\vec{v}_{\mathrm{p,LAB}}^{}$, is obtained by adding the velocity
of the compound nucleus in the laboratory system $\overrightarrow{v}_{\mathrm{CN},\mathrm{LAB}}^{\mathrm{}}$
and the velocity of the particle in the center of mass system $\overrightarrow{v}_{\mathrm{p},\mathrm{CM}}^{\mathrm{}}$: 

\begin{equation}
\overrightarrow{v}_{\mathrm{p},\mathrm{LAB}}^{\mathrm{}}=\overrightarrow{v}_{\mathrm{CN},\mathrm{LAB}}^{}+\overrightarrow{v}_{\mathrm{p,}\mathrm{CM}}^{}.\end{equation}

The transformation of the particle emission angle between the center
of mass m and laboratory systems is given by \\
\begin{equation}
\tan\theta_{\mathrm{p},\mathrm{LAB}}=\frac{\sin\mathrm{\theta}_{\mathrm{p,CM}}}{\cos\theta_{\mathrm{p,CM}}+x},\end{equation}
where \begin{equation}
x=\left[\frac{m_{\mathrm{a}}m_{\mathrm{b}}}{m_{\mathrm{A}}m_{\mathrm{B}}}\frac{E}{E_{}+Q_{\mathrm{}}}\right]^{1/2}.\end{equation}
Here $Q$ denotes the Q-value of the decay, $m_{\mathrm{a}}$, $m_{\mathrm{A}}$,
$m_{\mathrm{b}}$, $m_{\mathrm{B}}$ are the masses of initial and
final products, and $E$ the center of mass energy of the impinging
particle. 

An output file with the format shown in table \ref{tab:evaporoutput}
is produced by evapOR.

\begin{table}[H]
\begin{centering}
\begin{tabular}{|c|c|c|c|c|c|}
\hline 
 & \textbf{\small Index} & \textbf{\small Energy} & \textbf{\small vx} & \textbf{\small vy} & \textbf{\small vz}\tabularnewline
\hline
\hline 
\textbf{\small Start new event} & {\small \$} &  &  &  & \tabularnewline
\hline 
\textbf{\small Compound nucleus (Z,N)} & {\small 101 (18,39)} & {\small 63090} & {\small 0} & {\small 0} & {\small 1}\tabularnewline
\hline 
\textbf{\small Emitted particle ($\alpha$)} & {\small 7} & {\small 3615.4} & {\small -0.81693864} & {\small -0.13911782} & {\small 0.55969405}\tabularnewline
\hline 
\textbf{\small $\gamma$} & {\small 1} & {\small 2169.5} & {\small 0} & {\small 1} & {\small 0}\tabularnewline
\hline
\end{tabular}
\par\end{centering}

\centering{}\caption{\label{tab:evaporoutput}Example of the .aga output file produced
by evapOR. In the output shown here, the $\gamma$ ray is emitted
in the y direction. The parameters vx, vy, vz are the normalized $\left(\mid\overline{v}\mid=1\right)$
velocities in Cartesian coordinates in the laboratory system. The
indices 101, 7 and 1 denote the compound nucleus, an alpha particle
and a $\gamma$ ray respectively. Energy is the kinetic energy of
the particle in the laboratory system or the Doppler shifted $\gamma$-ray
energy, all given in keV. }

\end{table}

The event shown in table \ref{tab:evaporoutput} starts with the compound
nucleus. The listed $\gamma$ ray was not generated by evapOR but
was added artificially \cite{MarcinPalcz}. The energy of the $\gamma$
ray is Doppler shifted according to the non-relativistic Doppler formula
of eq. (\ref{eq:doppler2}).

\section{\label{sub:Geant4-&-Agata} Simulation of $\gamma$-ray detection}

The complex geometry of the AGATA detectors have been implemented
in the G\emph{\noun{eant}}4 simulation package \cite{agatacode,2003}.
In this package, one can choose between using a single HPGe crystal
or any number of triple clusters up to the full $4\pi$ AGATA array. 

One can either generate particles in G\emph{\noun{eant}}4 or get them
from an external event file. In the present simulations event files
from evapOR was used as input to \noun{Geant4}. The output of an ordinary
run using an input event file similar to the output shown in table
\ref{tab:evaporoutput}, is shown in table \ref{tab:agataoutput}.

\begin{table}[H]
\begin{centering}
\begin{tabular}{|c||c||c||c||c||c|}
\hline 
 & \textbf{\small Index} & \textbf{\small Energy | v/c} & \textbf{\small x} & \textbf{\small y} & \textbf{\small z}\tabularnewline
\hline
\hline 
\textbf{\small Start new event} & {\small -100} &  &  &  & \tabularnewline
\hline 
\textbf{\small Compound nucleus velocity} & {\small -101} & {\small v/c=0.05889} & {\small 0N} & {\small 0N} & {\small 1N}\tabularnewline
\hline 
\textbf{\small Position of CN} & {\small -102} &  & {\small 0} & {\small 0} & {\small 0}\tabularnewline
\hline 
\textbf{\small Emitted particle ($\alpha$)} & {\small -7} & {\small 3303.4} & {\small -0.04844N} & {\small -0.89261N} & {\small -0.44821N}\tabularnewline
\hline 
\textbf{\small Residual nucleus} & {\small -101} & {\small v/c=0.06147} & {\small 0.00175 N} & {\small 0.03233 N} & {\small 0.99948 N}\tabularnewline
\hline 
\textbf{\small Position of RN} & {\small -102} &  & {\small 0} & {\small 0} & {\small 0}\tabularnewline
\hline 
\textbf{\small Emitted particle $\gamma$} & {\small -1} & {\small 2172.300} & {\small 0N} & {\small 1N} & {\small 0N}\tabularnewline
\hline 
\textbf{\small Interaction point 1 (detector 2)} & {\small 2} & {\small 383.495} & {\small -0.373} & {\small 43.248} & {\small -0.426}\tabularnewline
\hline 
\textbf{\small Interaction point 2 (detector 2}) & {\small 2} & {\small 32.573} & {\small -1.477} & {\small 43.844} & {\small -1.851}\tabularnewline
\hline 
\textbf{\small Interaction point 3 }(\textbf{\small detector 2)} & {\small 2} & {\small 34.293} & {\small -7.208} & {\small 49.511} & {\small -2.958}\tabularnewline
\hline 
\textbf{\small RN after emitted $\gamma$ } & {\small -101} & {\small v/c=0.06147} & {\small 0.00175 N} & {\small 0.03133 N} & {\small 0.99951 N}\tabularnewline
\hline 
\textbf{\small Position of RN} & {\small -102} &  & {\small 0} & {\small 0} & {\small 0}\tabularnewline
\hline 
\textbf{\small RN after emitted $\gamma$ } & {\small -8} & {\small 66971.702} & {\small 0.00175 N} & {\small 0.03133 N} & {\small 0.99951 N}\tabularnewline
\hline
\end{tabular}
\par\end{centering}

\begin{centering}
\caption{\label{tab:agataoutput}Example of an output file of the AGATA G\emph{\noun{eant}}4
package. N denotes that the velocity of the residual nucleus is normalized
to 1. Energies are given in keV and positions in cm. RN stands for
the residual nucleus. }

\par\end{centering}

\end{table}

The events start (Index=-100) with the compound nucleus velocity divided
by the speed of light in vacuum, c, and its direction (-101), followed
by the position of the CN (-102), the evaporated particle (-7), and
the residual nucleus (-101, -102). The nucleus then emits the Doppler
shifted $\gamma$-ray (-1), which interacts three times in detector
number 2 (Index=2). After this the $v/c$ and the direction (-101)
and position (-102) of the residual nucleus is listed. Finally the
kinetic energy and the direction of the residual nucleus is given
(-8).

\section{Gamma-ray tracking }

The next step is to track the $\gamma$ rays \cite{Lopez-Martens_NIM_A533}.
In this work the MGT tracking program \cite{DinoBazacco} was used.
This tracking code uses the Compton scattering formula to determine
the path of the $\gamma$ ray in the detectors and gives the full
energy and the first interaction point of the incoming $\gamma$ ray.
The first interaction point provides the angle $\theta_{\mathrm{\gamma}}$
needed for the Doppler correction according to equation \ref{eq:doppler2}. 

In the planned AGATA commissioning experiment, no ancillary detectors
for the detection of the residual nuclei will be used, therefore the
velocity vectors of the residual nucleus will be unknown experimentally.
Instead all the residual nuclei are assumed to have the average velocity
and direction along the beam axis.

\chapter{Results and discussions}

\section{\label{sec:TALYS}\label{sub:Studied-reactions}TALYS}

In table \ref{tab:Proton-and-Deuteron} the studied reactions are
summarized. Following the requirements in chapter \ref{sub:Requirements},
most of the ion beams at LNL {[}appendix \ref{cha:Available-ion-beams}{]}
were excluded from further study. In most reactions, competition between
different reaction channels were found. The $\gamma$-ray energy was
also in many cases lower than required. The proton and deuteron capture
reactions had too small cross sections to be useful for the planned
AGATA commissioning experiment, hence only fusion evaporation reactions
were studied. 

\textbf{}%
\begin{table}[H]
\begin{centering}
\begin{tabular}{ll|ll}
\hline 
p - Reactions & Result & d - Reactions & Result\tabularnewline
\hline
 &  &  & \tabularnewline
$\mathrm{^{19}F+p}$  & {\footnotesize $\alpha$ channel dominates} & $^{27}\mathrm{Al+d}$ & {\footnotesize Competing p \& n channels}\tabularnewline
$^{63}\mathrm{Cu+p}$ & {\footnotesize p channel dominates} & $^{63}\mathrm{Cu+d}$ & {\footnotesize Competing n, p \& $\alpha$ channels}\tabularnewline
$^{81}\mathrm{Br+p}$ & {\footnotesize n ch. dominates, odd-even nucleus} & $^{107}\mathrm{Ag+d}$ & {\footnotesize Channels competing}\tabularnewline
$^{27}\mathrm{Al+p}$ & {\footnotesize $\alpha$ \& p channels competing} & $^{28}\mathrm{Si+d}$ & {\footnotesize p channel dominates}\tabularnewline
$^{31}\mathrm{P+p}$ & {\footnotesize p channel dominates} & \textbf{$^{37}\mathrm{Cl+d}$} & \textbf{\footnotesize n channel dominates }\tabularnewline
$^{35}\mathrm{Cl+p}$ & {\footnotesize p channel dominates} & \textbf{$^{51}\mathrm{V+d}$} & \textbf{\footnotesize n channel dominates}\tabularnewline
$^{58}\mathrm{Ni+p}$ & {\footnotesize p channel dominates} & $\mathrm{^{100}Mo+d}$ & {\footnotesize p channel dominates}\tabularnewline
$^{56}\mathrm{Fe+p}$ & {\footnotesize p channel dominates} & $^{94}\mathrm{Mo+d}$ & {\footnotesize p channel dominates}\tabularnewline
$^{28}\mathrm{Si+p}$ & {\footnotesize p channel dominates} & $\mathrm{^{81}Br+d}$ & {\footnotesize p \& n channels competing}\tabularnewline
$^{32}\mathrm{S+p}$ & {\footnotesize p channel dominates} &  & \tabularnewline
$\mathrm{^{33}S+p}$ & {\footnotesize p channel dominates} &  & \tabularnewline
$^{65}\mathrm{Cu+p}$ & {\footnotesize n ch. dominates, bad }$\gamma${\footnotesize{} energy} &  & \tabularnewline
 &  &  & \tabularnewline
\end{tabular}
\par\end{centering}

\textbf{\caption{\label{tab:Proton-and-Deuteron}Proton and deuteron reactions studied
by TALYS. }
}
\end{table}

\begin{figure}[H]
\begin{centering}
\includegraphics[scale=0.39]{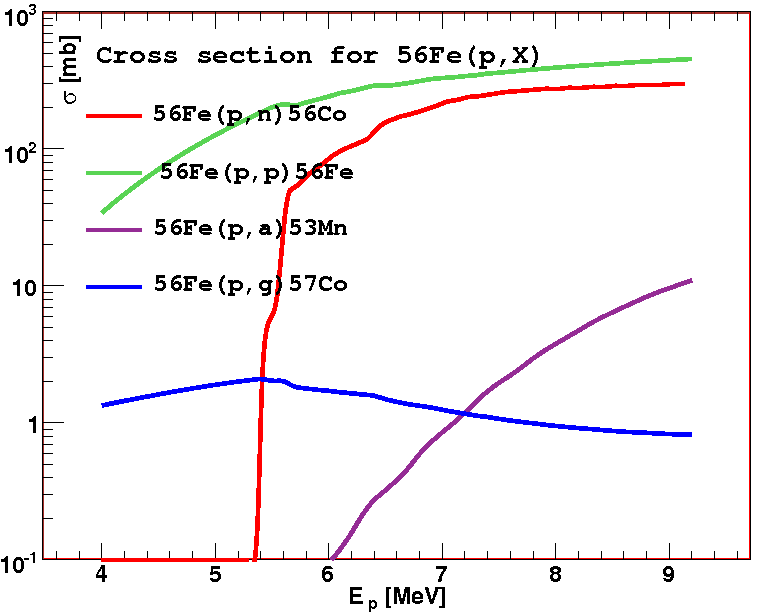}\includegraphics[scale=0.39]{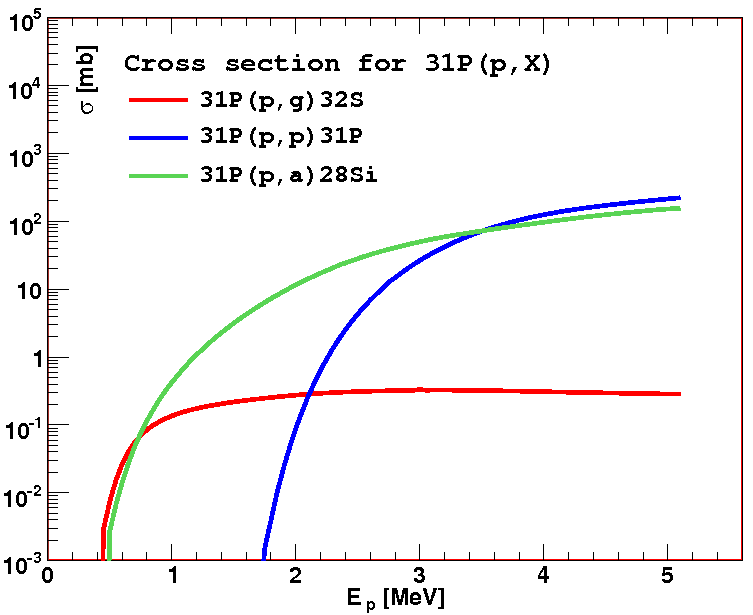}
\par\end{centering}

\caption{\label{fig:Example-two-(p,n)}Example of two (p,n) reactions, $^{56}\mathrm{Fe}(\mathrm{p,\mathit{X}})$
and $^{31}\mathrm{P}(\mathrm{p,\mathit{X}})$. (a=$\alpha$ particle,
g=$\gamma$ ray). The energy of the incident proton is given in the
laboratory system. }

\end{figure}
 In the left plot of figure \ref{fig:Example-two-(p,n)} the $\mathbf{\mathrm{^{56}Fe(p,p)^{56}Fe}}$
reaction channel is dominant. In the right plot of figure \ref{fig:Example-two-(p,n)},
the $\alpha$ emission dominates up to $\sim3$ MeV at which energy
proton emission starts to become important.

Two examples of deuteron induced reactions are presented in figure
\ref{fig:Example-two-(d,n)}. To the left, $^{28}\mathrm{Si(\mathrm{d},\mathit{X})}$,
where the proton channel dominates. In figure \ref{fig:Example-two-(d,n)},
$^{27}\mathrm{Al}(\mathrm{d},X)$ is shown, where there is a competition
between the neutron and proton channels .

\begin{figure}[H]
\begin{centering}
\includegraphics[scale=0.38]{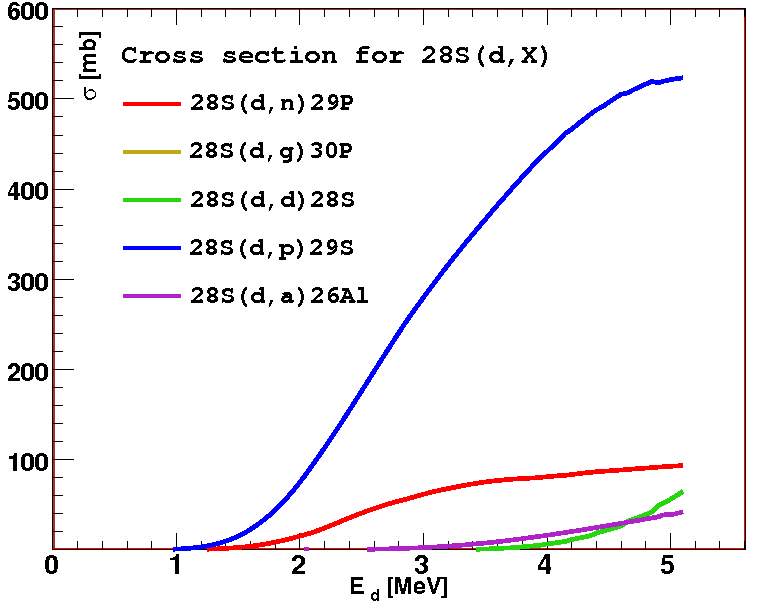}\includegraphics[scale=0.39]{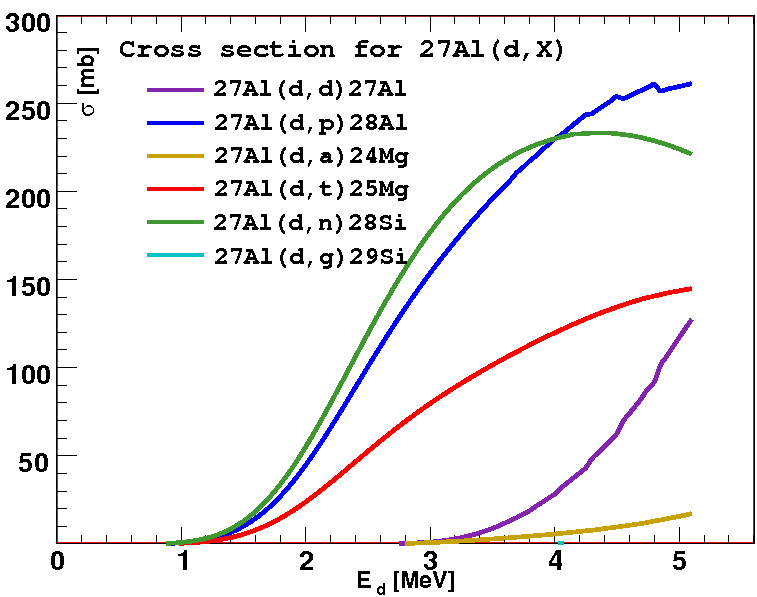}
\par\end{centering}

\caption{\label{fig:Example-two-(d,n)}Example of two (d,n) reactions, $^{28}\mathrm{Si(d,}X)$
and $^{27}\mathrm{Al(d,}X)$. The energy of the incident deuteron
is given in the laboratory system.}

\end{figure}

\begin{figure}[H]
\begin{centering}
\includegraphics[scale=0.39]{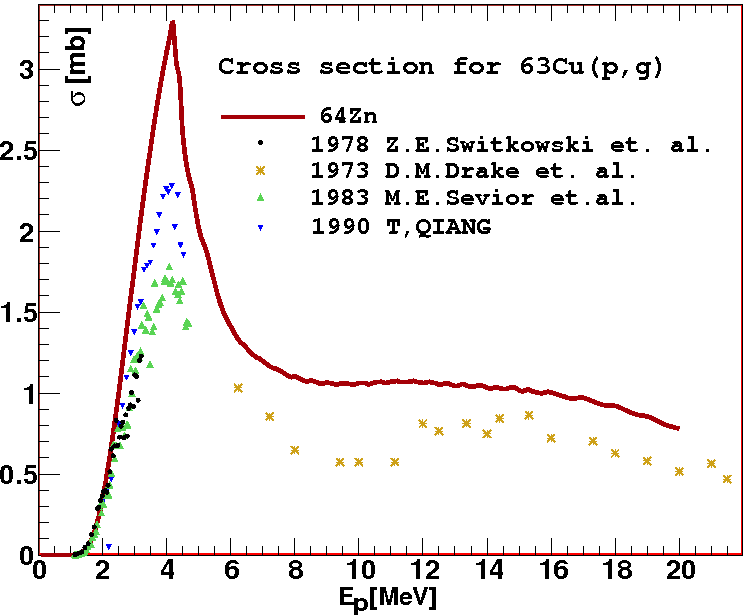}\includegraphics[scale=0.4]{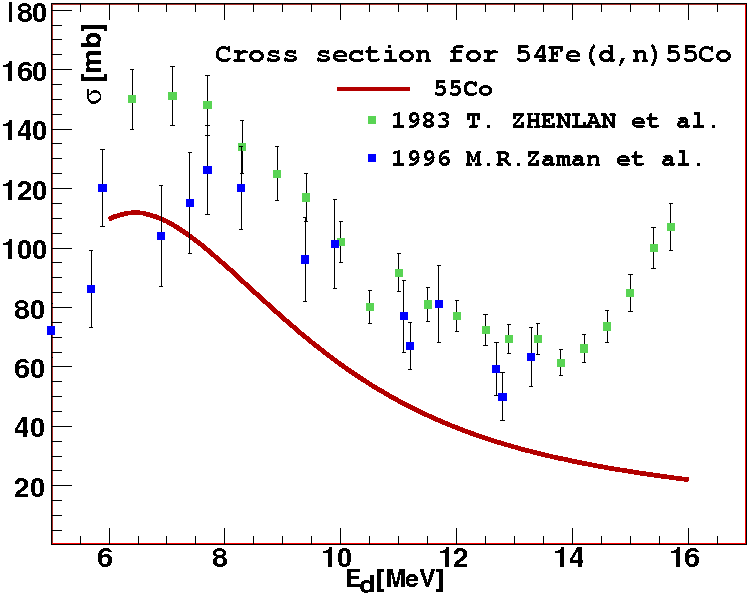}
\par\end{centering}

\caption{\label{fig:In-the-left-0}In the left figure the reaction $^{63}\mathrm{Cu(d,n)}$
is compared to measurements reported in ref. \cite{qiang,Sevior,drake,switkov}.
In the right figure $^{54}\mathrm{Fe(d,n)}$ is compared to experimental
results from ref. \cite{Zhenlan,zaman}. The incident proton and deuteron
energies are given in the laboratory system. }

\end{figure}

\begin{figure}[H]
\begin{centering}
\includegraphics[scale=0.4]{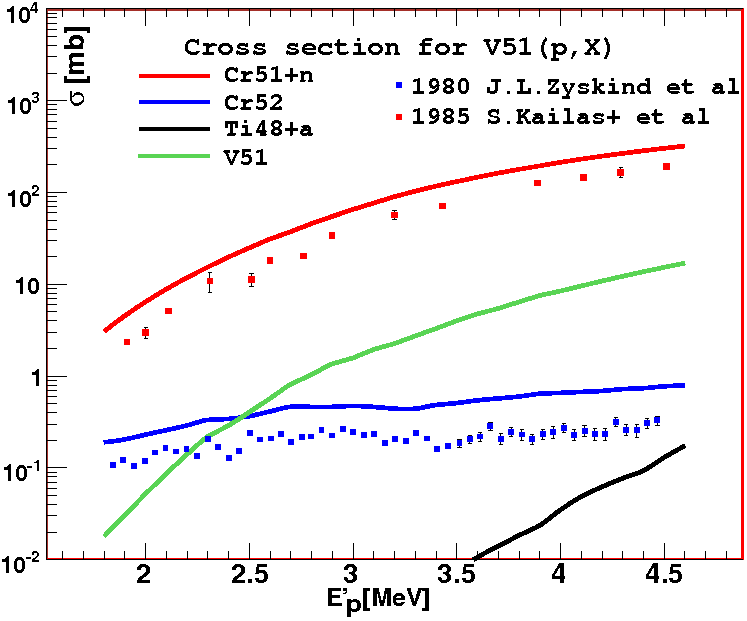}
\par\end{centering}

\caption{\label{expvan}The $^{51}\mathrm{V(p,n)}$ reaction compared to experimental
results from ref.  \cite{kailas,zyskind}. The energy of the incident
proton is given in the laboratory system. }

\end{figure}

Results obtained from the TALYS simulations were verified against
experimental data. Three examples of such comparisons are presented
in figures \ref{fig:In-the-left-0} and \ref{expvan}.

For the $\mathrm{^{63}Cu\left(p,\gamma\right)}$ reaction (figure
\ref{fig:In-the-left-0} left), four independent measurements were
plotted together with the calculation of TALYS. The differences are
small at low energies but reach a factor of 2 around the peak at $E_{\mathrm{p}}\approx4$
MeV. In the plot on the right hand side of figure \ref{fig:In-the-left-0},
the experimental values are close to the results of the TALYS calculation
for the $^{54}\mathrm{Fe\left(d,n\right)}$ reaction for energies
up to about $14$ MeV. At higher energies the measurements of ref.
\cite{Zhenlan} differ from the TALYS calculations. In figure \ref{expvan}
the reaction $\mathrm{^{51}V\left(p,X\right)}$ is simulated and compared
with measurements from ref. \cite{kailas,zyskind}. Both the neutron
and $\gamma$-ray emission channels are close to the results obtained
from measurements.

\subsection{\label{sub:Adequate-reactions-for}Selected reactions }

\begin{figure}[H]
\begin{centering}
\includegraphics[scale=0.49]{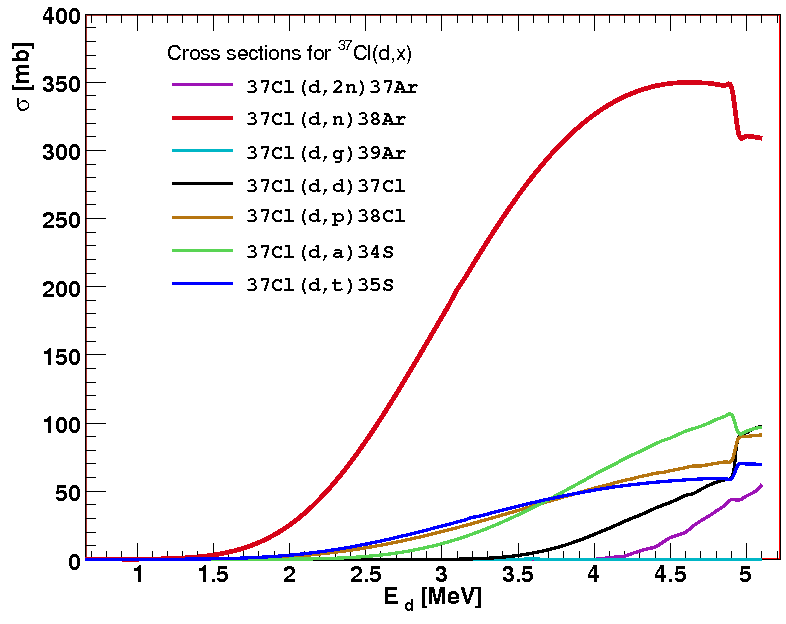}
\par\end{centering}

\begin{centering}
\includegraphics[scale=0.5]{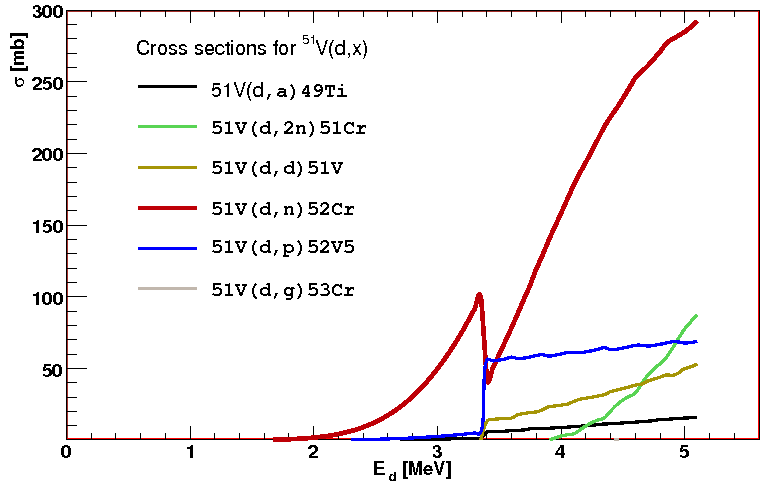}
\par\end{centering}

\caption{\label{fig:TALYS-cross-section}TALYS cross-section calculations for
$\mathrm{^{37}Cl\left(d,n\right)Ar}$ and $\mathrm{^{51}V\left(d,n\right)Cr}$.
The incident deuteron energies are given in the laboratory system.}

\end{figure}

In table \ref{tab:Proton-and-Deuteron} two reactions were shown with
a dominating neutron emission channel. These are the reactions that
were further studied. Both reactions leads to an even-even residual
nucleus, and they have appropriate life times. The first reaction,
$\mathrm{^{37}Cl\left(d,n\right){}^{38}Ar}$, leads to $^{38}\mathrm{Ar}$
which has a first excited state $2^{+}$ at 2168 keV. The high $\gamma$-ray
energy which decays to the $0^{+}$ ground state is very good for
the Doppler-effect measurements. The cross-section calculation from
TALYS is shown in the upper plot of figure \ref{fig:TALYS-cross-section}.
The reaction $\mathrm{^{37}Cl(d,n)^{38}Ar}$ is clearly dominating. 

The second reaction, $\mathrm{^{51}V\left(d,n\right){}^{52}Cr},$
is for the same reason a suitable reaction. The first excited state
has an excitation energy of 1434 keV. The calculated cross section
for this reaction is shown in the lower plot of figure \ref{fig:TALYS-cross-section}.

 The cross section of the reaction $^{37}\mathrm{\mathrm{Cl}\left(d,n\right)}\mathrm{{}^{38}Ar}$
reaches a maximum of about 350 mb at a deuteron energy of about 4
MeV. The cross section of the $\mathrm{^{51}V\left(d,n\right){}^{52}Cr}$
reaction has a local maximum at about $\mathrm{E_{d,LAB}=3}.4$ MeV
and it increases at higher energies without competition with other
channels. Based on the TALYS results, both reactions, $\mathrm{d+^{37}Cl}$
and $\mathrm{d+^{51}V}$, are suitable for further studies. The energies
chosen are $E_{d}=3.8$MeV and $E_{d}=3.2$MeV for the $\mathrm{d+^{37}}\mathrm{Cl}$
and $\mathrm{d}+^{51}\mathrm{V}$ reactions respectively. In inverse
kinematics reactions thus correspond to $E_{37_{\mathrm{Cl}}}=70.0$
MeV and $E_{51_{\mathrm{V}}}=81.6$ MeV, respectively.

\subsection{\label{sub:Compared-performance-test}Comparison between evapOR and
TALYS}

\begin{figure}[H]
\begin{centering}
\includegraphics[angle=90,scale=0.7]{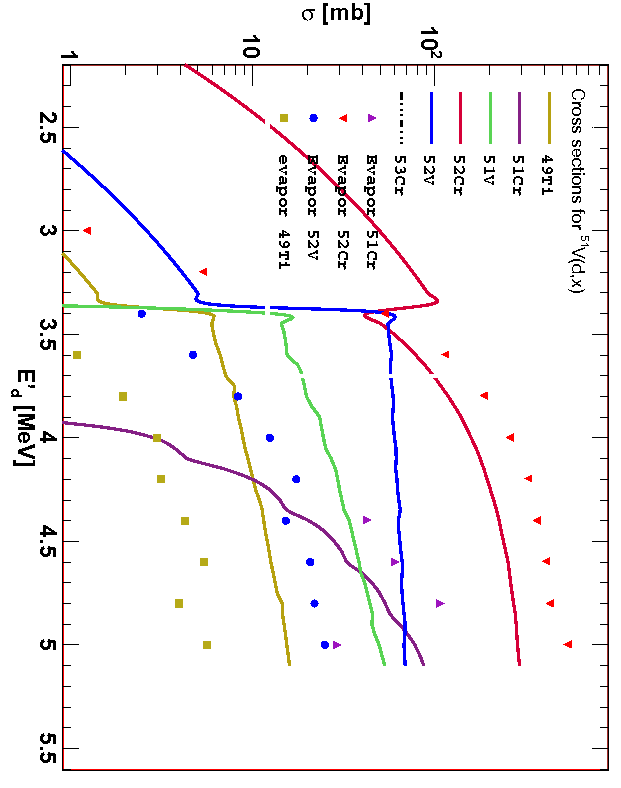}
\par\end{centering}

\caption{\label{fig:Results-obtained-from}Comparison of cross sections obtained
by TALYS (lines), and evapOR (symbols) for the $\mathrm{d+^{51}V}$
reaction. The incident deuteron energy is given in the laboratory
system.}

\end{figure}

To verify the calculations, a comparison between TALYS and evapOR
was made. The results for the $\mathrm{d+^{51}V}$ reaction are presented
in figure \ref{fig:Results-obtained-from}.

 As seen in figure \ref{fig:Results-obtained-from}, large discrepancies
were obtained. Below $E_{\mathrm{}}\approx3.4\mathrm{\; MeV}$ the
TALYS and evapOR results are closer. Above 3.4 MeV the cross sections
are similar in particular for the $\mathrm{^{51}V\left(d,n\right)^{52}Cr}$
reaction.

\section{\label{sec:Results-target-studies}Studies of target effects}

In the proposed commissioning experiment, an inverse kinematics reaction
will be used. This means that for the two selected reactions, $\mathrm{d+^{37}Cl}$
and $\mathrm{d+^{51}V}$, deuterium would be the target nucleus. One
possible target is to use a thin plastic foil (e.g. polyethylene)
enriched in deuterium. Such foils do, however, only allow very weak
beam currents without the foil getting destroyed (melting). Instead
it is proposed to use a deuterated Ti foil which can contain about
1 deuterium atom per titanium atom\cite{A.Gadea}. The chosen target
thickness is 200 $\mathrm{\mu m}$/$\mathrm{cm^{^{2}}}$. No deuterated
titanium compound is available in TRIM therefore pure titanium was
chosen as target in the simulation. The difference in energy straggling
due to this is expected to be negigible. No nuclear reactions will
occur between the $\mathrm{^{37}Cl}$ or $^{\mathrm{51}}\mathrm{V}$
beams and the Ti nuclei of the target (the energies are well below
the Coloumb barrier).

\subsection{Energy straggling}

\begin{figure}[H]
\begin{centering}
\includegraphics[scale=0.8]{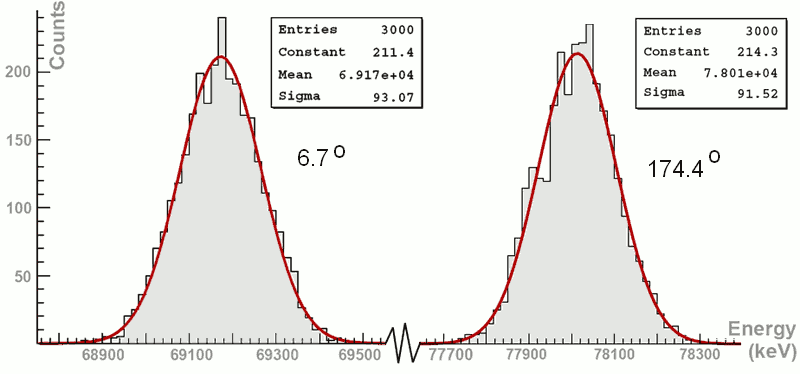}
\par\end{centering}

\caption{\label{fig:TRIM-calculation-of}TRIM calculations of energy straggling
of $^{52}\mathrm{Cr}$ ions in a 200 $\mu\mathrm{m}$/$\mathrm{cm^{^{2}}}$thick
titanium target. In the figure, the kinetic energy distributions of
the residual nuclei when they exit the target material are plotted
for the case when neutrons are emitted at the angles $\theta_{\mathrm{n,CM}}^{\mathrm{}}=174.4^{\circ}$
(right) and $\theta_{\mathrm{n,CM}}^{\mathrm{}}=6.7^{\circ}$ (left)
corresponds to initial recoil energy of $E_{\mathrm{RN1}}=82.4\:\mathrm{MeV}$
and $E_{\mathrm{RN2}}=73.6\:\mathrm{MeV}$, respectively. }

\end{figure}

The target studies were based on the two effects presented in section
\ref{sub:Sources-of-target}, due to the neutron distribution and
the position of the reaction. Here only results of the study of the
reaction $\mathrm{d+^{51}V}$ are presented. The results of the target
studies for the $\mathrm{d+^{37}Cl}$ reaction are very similar. In
order to study the effect of different neutron emission angles, two
events with relatively high and low residual nucleus ($^{52}\mathrm{Cr}$)
kinetic energies were chosen from the evapOR simulation. The energies
were $E_{\mathrm{RN1}}=82.4$ MeV and $E_{\mathrm{RN2}}=73.6$ MeV
in the laboratory system. The difference is due to the neutron emission
angle: the event with energy $E_{\mathrm{RN}1}$ had a large neutron
emission angle, $\theta_{\mathrm{n,CM}}^{\mathrm{}}=174.4^{\circ}$
while the event with $E_{\mathrm{RN2}}$ had a small angle, $\theta_{\mathrm{n},\mathrm{CM}}^{}=6.7{}^{\circ}$.
The energies $E_{\mathrm{RN1}}$ and $E_{\mathrm{RN2}}$ were used
as input to TRIM. The energy losses of the residual nuclei are shown
in figure \ref{fig:TRIM-calculation-of}. %
 For the two cases the average energy losses are 

\[
E_{\mathrm{RN1}}^{\mathrm{loss}}=82.40\;\mathrm{MeV}-78.01\;\mathrm{MeV}=4.390\;\mathrm{MeV},\]
which gives the velocity

\begin{equation}
\frac{\triangle v_{RN1}^{\mathrm{loss}}}{c}=\sqrt{\frac{2}{m_{RN1}c^{2}}}\left(\sqrt{E_{RN1,in}}-\sqrt{E_{RN1,out}}\right)=1.5760\cdot10^{-3}.\end{equation}
And

\[
E_{\mathrm{RN2}}^{\mathrm{loss}}=73.60\;\mathrm{MeV}-69.17\;\mathrm{MeV}=4.430\;\mathrm{MeV,}\]
which gives the velocity \begin{equation}
\frac{\triangle v_{RN2}^{\mathrm{loss}}}{c}=\sqrt{\frac{2}{m_{RN2}c^{2}}}\left(\sqrt{E_{RN,in}}-\sqrt{E_{RN,out}}\right)=1.6858\cdot10^{-3}.\end{equation}

Next, the effects of this energy-loss difference on the Doppler shifts
were studied. In the case of the $^{52}\mathrm{Cr}$ residual nucleus,
the $\gamma$-ray energy from the $2^{+}$ excited state to the $0^{+}$
ground state is 1434 keV. For detector position at $90^{\mathrm{o}}$relative
to the incoming beam and a distance of 15 cm from the target to the
front face of the detector, which was chosen for these simulations
(see section \ref{sec:Spectra}), the smallest angle for $\gamma$-ray
detection, which corresponds to the largest Doppler shift, is about
$70^{\mathrm{o}}$. Thus the difference in $\gamma$ ray energy becomes\begin{equation}
\Delta E_{\gamma}=E_{\gamma_{0}}\cdot\frac{\triangle v_{RN2}^{\mathrm{loss}}-\triangle v_{RN1}^{\mathrm{loss}}}{c}\cos\theta_{\gamma}=0.054\mathrm{keV,}\end{equation}
which is a small effect. The contribution from the Doppler shift from
this studied effect is hence negligible. 

\begin{figure}[H]
\begin{centering}
\includegraphics[angle=90,scale=0.5]{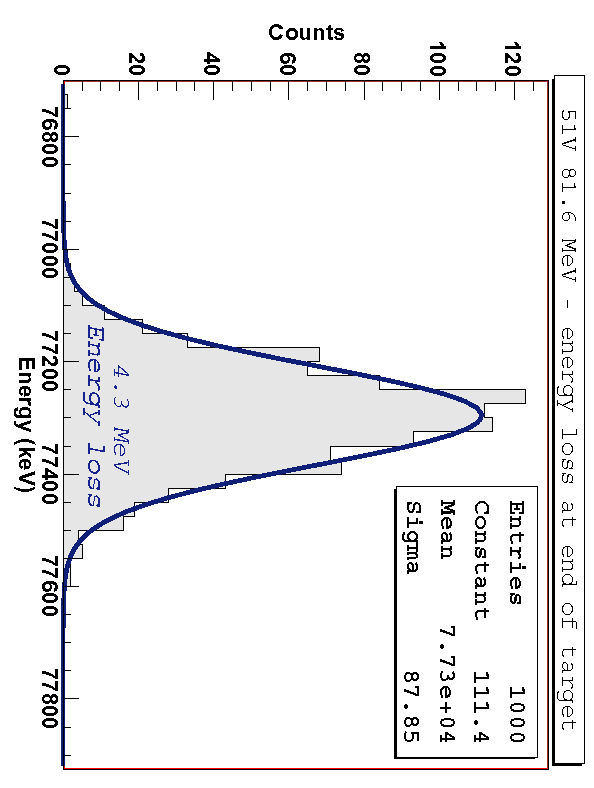}
\par\end{centering}

\caption{\label{fig:The-reaction-at}Energy distribution of $^{51}\mathrm{V}$
residual nuclei for the case that the reaction occurs at the end of
the target. The energy of the incident $^{51}\mathrm{V}$ ion beam
is 81.6 MeV. The target was a $200$ $\mu\mathrm{g/cm^{2}}$thick
Ti foil. }

\end{figure}

The second target effect concerns the position of the reaction. A
new TRIM calculation was performed for $^{51}\mathrm{V}$ ions with
an energy of $E=81.6$ MeV. The energy loss of the $^{51}\mathrm{V}$
beam in the target is shown in figure \ref{fig:The-reaction-at}.
 The energy loss of the beam through the target is,

\[
E_{^{51}\mathrm{V}}^{\mathrm{loss}}=81.6-77.3=4.3\;\mathrm{MeV}.\]
From the previous study of $^{52}\mathrm{Cr,}$ the average of $E_{\mathrm{RN1}}$
and $E_{\mathrm{RN2}}$ gives the average energy loss in the case
the reactions occur at the beginning of target:

\[
E_{^{52}\mathrm{Cr}}^{\mathrm{loss}}=\frac{E_{RN1}+E_{RN2}}{2}=4.41\;\mathrm{MeV.}\]
The velocity of the $^{52}\mathrm{Cr}$ residual nucleus for the case
of a reaction at the end of the target is 

\begin{equation}
\frac{v}{c}=0.0571,\end{equation}
and

\begin{equation}
\frac{v}{c}=0.0570,\end{equation}
if the reaction occurs in the beginning of the target. Note that it
is assumed that the $\gamma$ rays are emitted after the residual
nucleus has exited the target (see section ). The difference in velocities
is small and gives a negligible effect on the Doppler shifts at $\theta_{\gamma}=70^{\mathrm{o}}$:\begin{equation}
\Delta E_{\gamma}=E_{\gamma_{0}}\cdot\frac{\Delta v}{c}\cos\theta_{\gamma}=0.05\mathrm{keV}.\end{equation}
Hence, the position of the reaction in the target can be neglected. 

A similar study was done for the $\mathrm{d({}^{37}Cl,n)}$ reaction
showing a negligible effect on the Doppler shifts. The average energy
loss for the $\mathrm{d({}^{37}Cl,n)}$ reaction was

\begin{equation}
E_{^{37}\mathrm{Cl}}^{\mathrm{loss}}=3.50\:\mathrm{MeV}.\label{eq:resulttarget}\end{equation}

\subsection{Angular straggling}

\begin{figure}[H]
\begin{centering}
\includegraphics[scale=0.5]{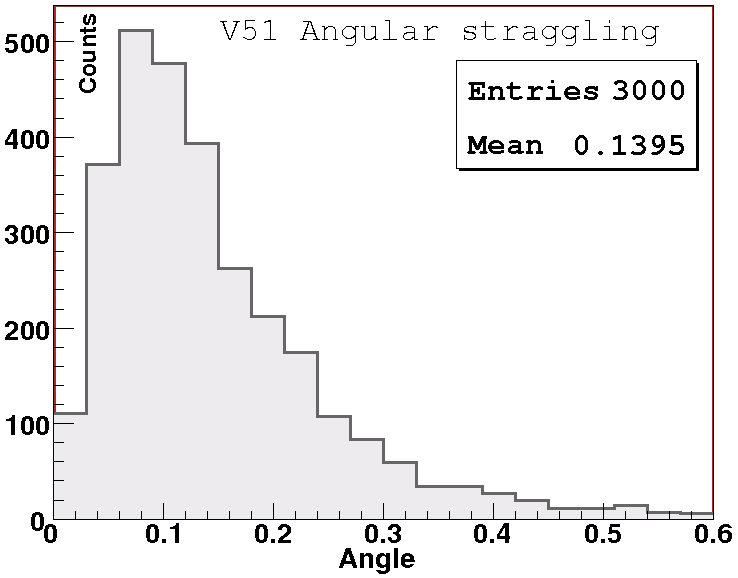}
\par\end{centering}

\caption{\label{fig:Angular-deflection}Angular straggling of 81.6 MeV $^{51}\mathrm{V}$
ions in a $200$ $\mu\mathrm{g/cm^{2}}$ thick Ti target.}

\end{figure}

In figure \ref{fig:Angular-deflection}, the angular deviation from
the beam direction of the $^{51}\mathrm{V}$ ions after passing through
the $200$ $\mu\mathrm{g/cm^{2}}$ thick Ti target is shown. %
 The angular distribution of the ions due to straggling has a maximum
at $0.1^{\circ}$. This is much smaller than the maximum of the angular
distribution due to the neutron emission, see fig. \ref{fig:angular effect doppler}.
Thus, the angular straggling in the target can be neglected.

\subsection{Life-time effects}

The compound nucleus has a very short life time compared to the time
it takes for it to pass through the target. To prevent the $\gamma$
rays to be emitted while the residual nucleus is still in the target,
it is required that the life time are long enough (see section \ref{sub:Requirements}).
If the $\gamma$ rays are emitted while the residual nucleus still
is located inside the target a too large spread of Doppler shift will
be the result. The average flight distances of the residual nuclei
before emitting the $\gamma$ rays were studied. For the reaction
$\mathrm{d({}^{51}V,n)^{52}\mathrm{Cr}}$ the average flight distance
$x$, before the emission of the $2^{+}\rightarrow0^{+}$ $\gamma$-ray
in $^{52}\mathrm{Cr}$ is 

\begin{equation}
x\approx1.0099\mathrm{ps}\cdot0.0548\mathrm{c}=1.15\cdot10^{-5}\mathrm{m}=16.5\mathrm{\mu m},\end{equation}
which is an order of magnitude larger than the target thickness $\approx0.5\mu m$.
The life-time of the $2^{+}\rightarrow0^{+}$ transition in $^{38}\mathrm{Ar}$
is 0.578 ps and gives

\begin{equation}
x\approx0.578\mathrm{ps}\cdot0.0589\mathrm{c}=10.0\mathrm{\mu m},\end{equation}
which is also large compared to the target thickness. Thus the photons
will be emitted outside the target material in both reactions. 

The effective life time of the transitions of interest must also be
shorter than the time it takes for the residual nuclei to travel a
distance $\lesssim1\mathrm{mm}$. This effect was not studied in the
present work.

\subsection{\label{sub:Experimental-detemination-of}Experimental determination
of the intrinsic FWHM }

\begin{figure}[H]
\begin{centering}
\includegraphics[angle=90,scale=0.5]{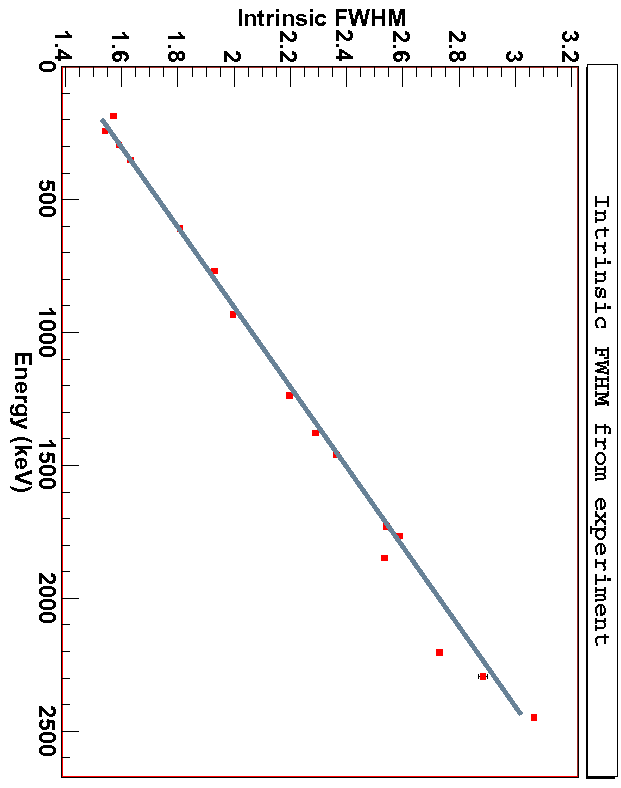}
\par\end{centering}

\caption{\label{fig:Intrinsic-FHWM-as}Intrinsic FWHM, $W_{\mathrm{i}}$, as
a function of $\gamma$-ray energy for AGATA crystal C002 \cite{A.Wiens}.}

\end{figure}

The intrinsic part of the FWHM, $W_{\mathrm{i}}$, given in equation
\ref{eq:310} is unique for every detector type. This value is used
as input to the tracking program MGT. Data from a measurement of the
core segment of AGATA crystal number C002 using a $^{226}\mathrm{Ra}$
source \cite{A.Wiens} was used to obtain the energy dependence of
$W_{\mathrm{i}}$ for a typical AGATA detector. A straight line was
fitted to the data points, see fig. \ref{fig:Intrinsic-FHWM-as}. 

 Using the fitted straight line of fig. \ref{fig:Intrinsic-FHWM-as},
the intrinsic FWHM of the peaks corresponding to the 1434keV and 2168keV
$2^{+}\rightarrow0^{+}$ transitions in $^{52}Cr$ and $^{38}Ar$
are 2.30keV and 2.83keV, respectively.

\section{\label{sec:evapOR}evapOR }

The studies of the target effect described in section \ref{sec:Results-target-studies},
yielded an average energy loss in the target of 4.3MeV and 3.5MeV
for the $^{51}\mathrm{V}$ and $^{37}\mathrm{Cl}$ induced reactions,
respectively. The laboratory energies selected for the $\mathrm{d({}^{51}V,n)^{52}Cr}$
and $\mathrm{d({}^{37}Cl,n){}^{38}Ar}$ reactions were $77.0\:\mathrm{MeV}$
and $66.5\mathrm{\: MeV}$, respectively.

\subsection{Energy and angular distributions}

\begin{figure}[H]
\includegraphics[scale=0.42]{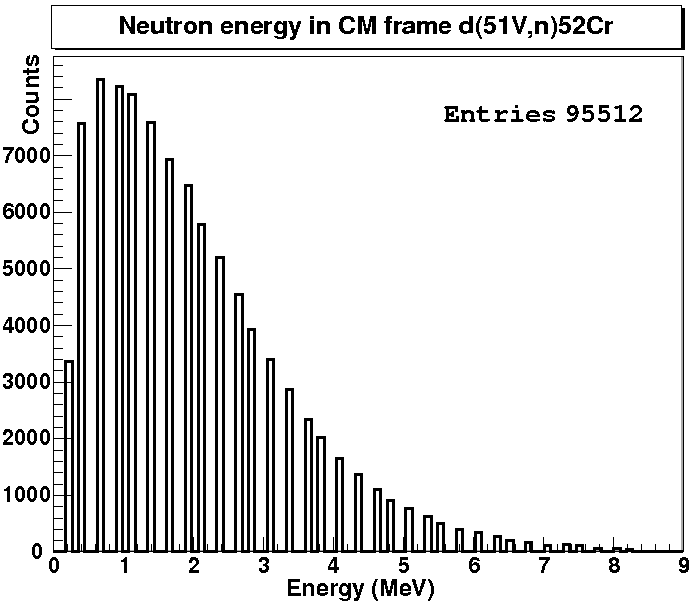}\includegraphics[scale=0.42]{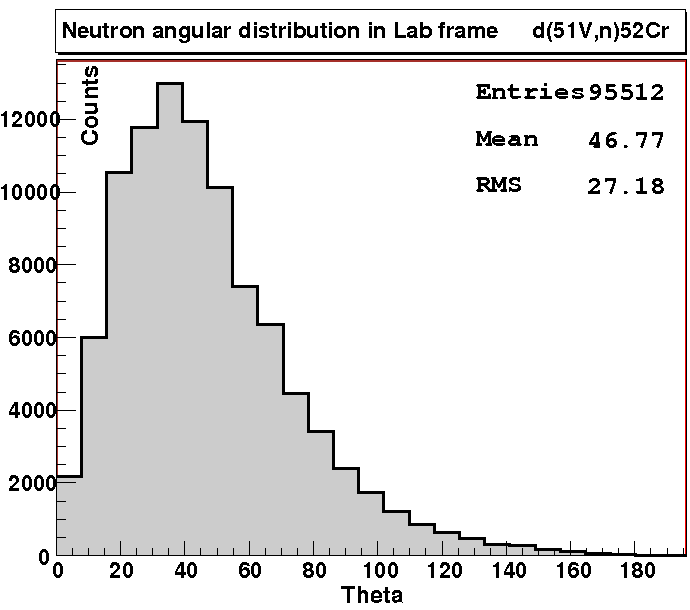}

\centering{}\caption{\label{fig:Ecm}The left panel shows the center of mass neutron energy
$E_{\mathrm{CM}}^{\mathrm{n}}$ and the right panel shows the angular
distribution of the neutrons in the laboratory system, $\theta_{\mathrm{LAB}}^{\mathrm{n}}$
for the reaction $\mathrm{d\left(^{51}V,n\right)^{52}Cr}$ at $\mathrm{\mathrm{\mathit{E}}\left(^{51}V\right)=77MeV}$. }

\end{figure}

Figure \ref{fig:Ecm} shows the energy and angular distribution of
the neutrons as obtained by evapOR.

 The energy distribution (left hand figure) is Maxwellian as expected
from equation \ref{eq:WW}. The distribution has a maximum at about
1MeV. According to the plot on the right hand side of figure \ref{fig:Ecm},
the neutrons have a large emission angle in the laboratory frame.
The centroid of the distribution is around $40^{\circ}$. 

\begin{figure}[H]
\centering{}\includegraphics[scale=0.45]{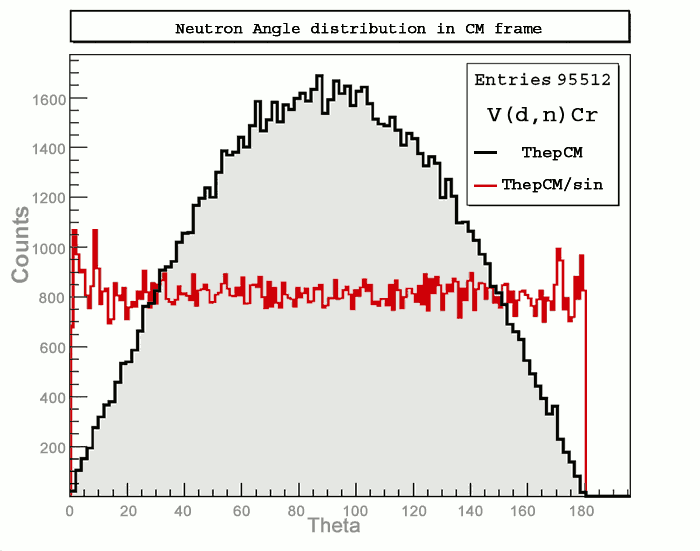}\caption{\label{fig:sintheta}Angular distribution of neutrons in the center
of mass system for the reaction $\mathrm{d({}^{51}V,n)^{52}Cr}$ at
$\mathrm{E\left(^{51}V\right)=77MeV}$. The black histogram shows
the number of neutrons emitted per $\theta_{CM}$ value. The red histogram
is the black histogram divided by $\sin\left(\theta_{CM}\right).$The
histograms are arbitrarily normalized to each other. }

\end{figure}

Figure \ref{fig:sintheta} shows the angular distribution of the neutrons
in the center of mass system. 

\begin{figure}[H]
\includegraphics[scale=0.44]{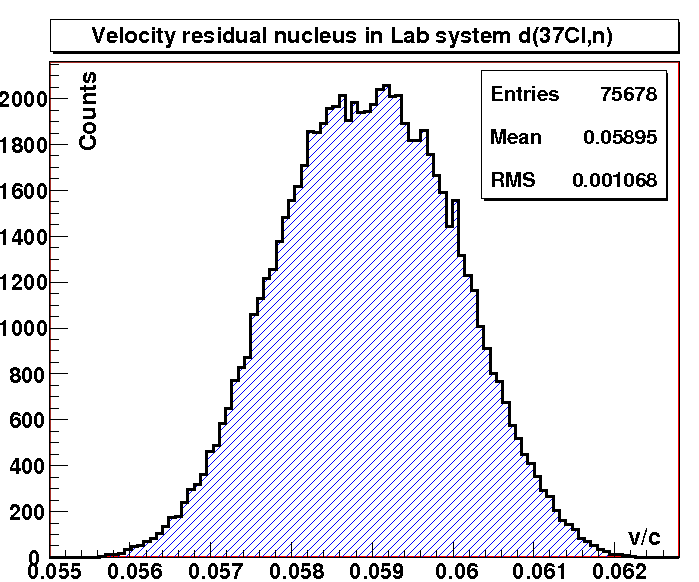}\includegraphics[scale=0.43]{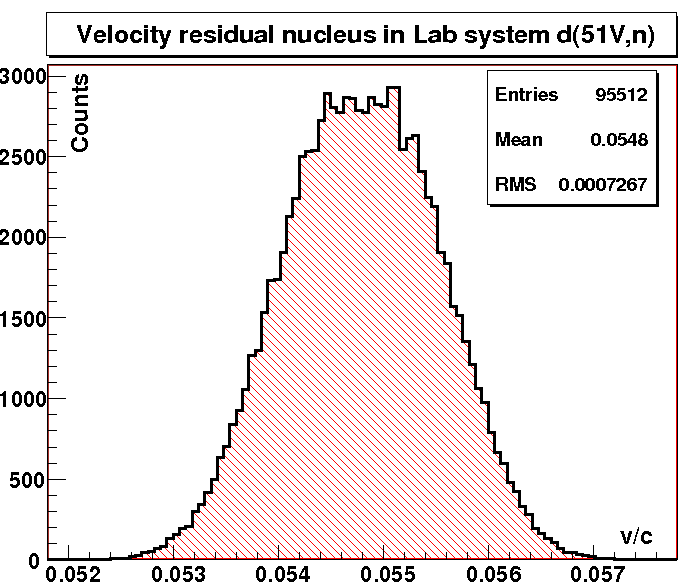}

\centering{}\caption{\label{fig:Velocity-distributions-for}Velocity distributions of the
residual nuclei for the $\mathrm{d\left(^{37}Cl,n\right)^{38}Ar}$
(left) and $\mathrm{d\left(^{51}V,n\right)^{52}Cr}$ (right) reactions
at $\mathrm{E\left(^{37}Cl\right)=66.5MeV}$ and $\mathrm{E\left(^{51}V\right)=77MeV}$,
respectively.}

\end{figure}

\begin{figure}[H]
\begin{centering}
\includegraphics[scale=0.42]{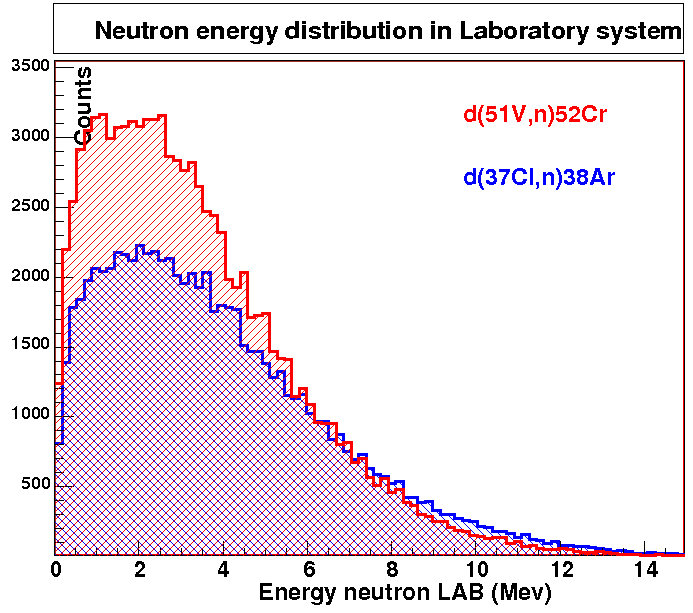}\includegraphics[scale=0.42]{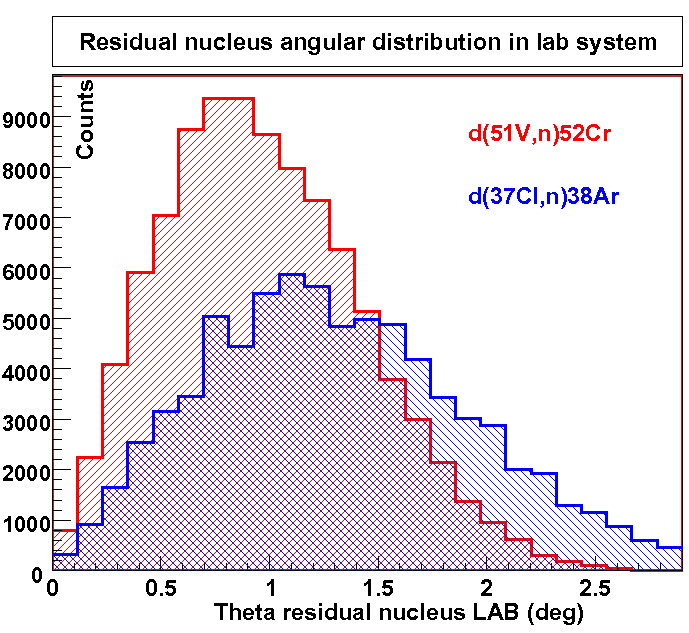}
\par\end{centering}

\centering{}\caption{\label{fig:angular effect doppler}The left panel shows the energy
distributions of the neutrons and the right panel the angular distribution
of the residual nuclei in the laboratory system.}

\end{figure}

Figure \ref{fig:Velocity-distributions-for} shows the velocity distribution
of the residual nuclei in both reactions. The neutron energy distributions
in the laboratory frame are shown in the left panel of figure \ref{fig:angular effect doppler}.
The $\mathrm{d({}^{51}V,n)^{52}Cr}$ reaction contains more neutron
events and thus the area under the red histogram is larger. The right
panel of figure \ref{fig:angular effect doppler} shows the angular
distributions of the residual nuclei. There is an evident shift of
the angular distribution, implying a smaller mean deflection of the
$^{52}\mathrm{Cr}$ nuclei compared to the $^{38}\mathrm{Ar}$ nuclei.
This difference in angular deflection plays a major role in the Doppler
shift of the $\gamma$-ray energies.

\section{\label{sec:Spectra}Simulated $\gamma$-ray spectra }

\subsection{Graphical illustration of the \noun{GEANT4} simulation}

\begin{figure}[H]
\centering{}\includegraphics[scale=0.45]{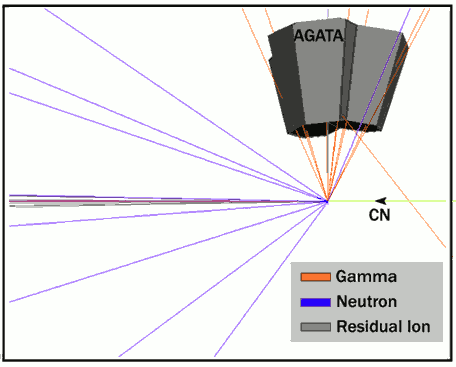}\caption{\label{fig:The-cluster-used}A graphical illustration of a \noun{Geant4}
simulation in which a compound nucleus emits neutrons and $\gamma$
rays, which are impinging on an AGATA triple cluster detector. See
text for further details. }

\end{figure}

In figure \ref{fig:The-cluster-used} a graphical illustration of
the \noun{Geant4} simulation is shown. Ten events of the reaction
$\mathrm{^{51}V\left(77\: MeV\right)+d}$ were generated by evapOR.
The direction of motion of the incoming $^{51}\mathrm{V}$ beam is
from right to left in the figure. For the plot only events belonging
to the reaction channel $\mathrm{d({}^{51}V,n)^{52}Cr}$ were selected.
Each $\mathrm{^{52}Cr}$ residual nucleus emits one $\gamma$ ray,
with an energy of 1434 keV, corresponding to the $2^{+}\rightarrow0^{+}$
transition in $^{52}\mathrm{Cr}$. %
 The $\gamma$ rays were forced to be emitted isotropically into a
cone with an opening angle of $20^{\mathrm{o}}$, centered at $90^{\mathrm{o}}$
relative to direction of incoming beam. An AGATA triple cluster detector
was placed at that angle, and covered a solid angle which was smaller
than the solid angle of the $\gamma$-ray emission cone. The distance
between the $\gamma$-ray source and the front of the HPGe crystals
was 15 cm in the figure.

Except if otherwise noted, the setup shown in fig. \ref{fig:The-cluster-used},
with the angle and distance given above, was used in the simulations
presented in the following subsections.

\subsection{Gamma-ray spectrum produced by summing}

As a first simple test, a $\gamma$-ray spectrum was created by summing
for each event the total energy deposited in the detector. Such a
spectrum is shown in fig. \ref{fig:nomgt.Spectra-for-^{38}Ar} for
the 2168 keV $\gamma$ ray emitted in the $\mathrm{d({}^{37}Cl,n){}^{38}Ar}$
reaction at $E_{\mathrm{LAB}}\left(^{37}\mathrm{Cl}\right)=66.5\:\mathrm{MeV}$.
The $\gamma$ rays were emitted isotropically into a $90^{\mathrm{o}}\pm0.5^{\mathrm{o}}$
cone. The intrinsic energy resolution was set to $0$ ($W_{\mathrm{i}}=0$
keV) and no Doppler corrections were applied. The full-energy peak,
the single- and double-escape peaks, and Compton background are clearly
visible in the spectrum.

\begin{figure}[H]
\begin{centering}
\includegraphics[scale=0.5]{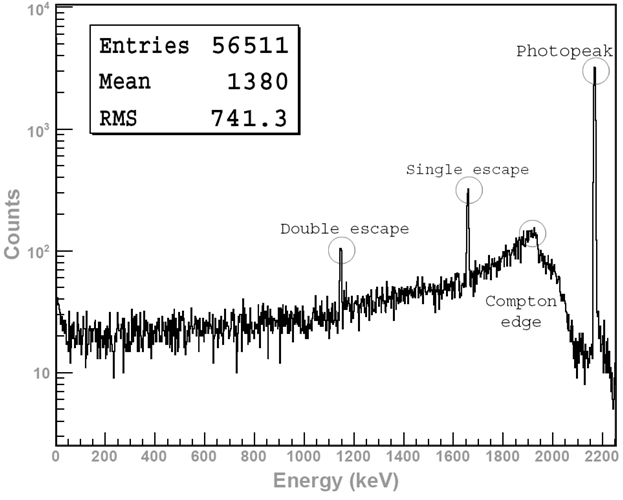}
\par\end{centering}

\caption{\label{fig:nomgt.Spectra-for-^{38}Ar}A $\gamma$-ray spectrum obtained
by summing the total energy absorbed in the AGATA triple cluster detector.
A total of $10^{5}$ evapOR events of the reaction $^{37}\mathrm{Cl}\left(66.5\;\mathrm{MeV}\right)+\mathrm{d}$
were generated. See text for further details. }

\end{figure}

\subsection{\label{sub:MGT-tracking-test.}Gamma-ray spectra produced by tracking
and Doppler correction}

Tracked $\gamma$-ray energy spectra were created by feeding the output
of the \noun{Geant4} simulation into the $\textsc{MGT}$ tracking
program. The $\textsc{MGT}$ smearing parameter, which emulates a
finite interaction position resolution in the HPGe crystals, was used
as one of the parameters in the simulations. 

The $\gamma$-ray emission angle, which is needed by the Doppler correction
procedure, was calculated by assuming that the $\gamma$ rays were
emitted from the center of AGATA and detected at the first interaction
point, which was obtained from $\textsc{MGT}$. For the Doppler correction
the velocity vector of the residual nucleus is also needed. Except
were otherwise noted, an average value of the velocity vector was
used, namely an angle of $0^{\mathrm{o}}$ (motion parallel to the
beam) and the average velocity given by the evapOR simulation. The
use of such average values is common in fusion-evaporation reactions,
when no ancillary detector is available for the event-by-event detection
of the velocity vector of the residual nuclei.

\begin{figure}[H]
\includegraphics[scale=0.42]{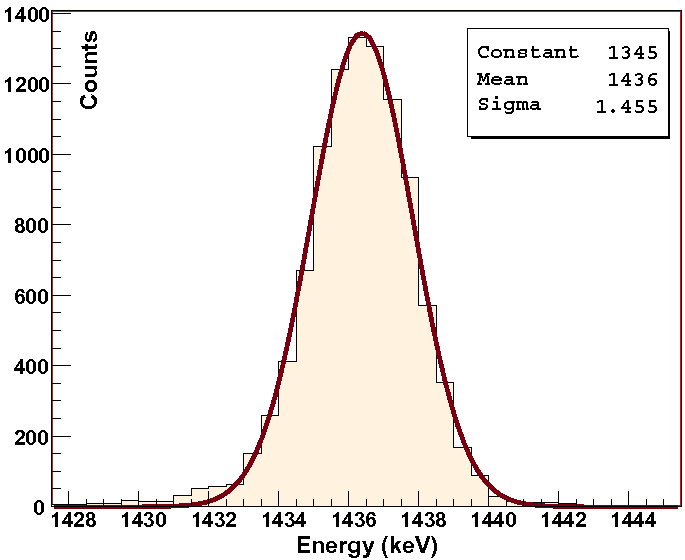}\includegraphics[scale=0.42]{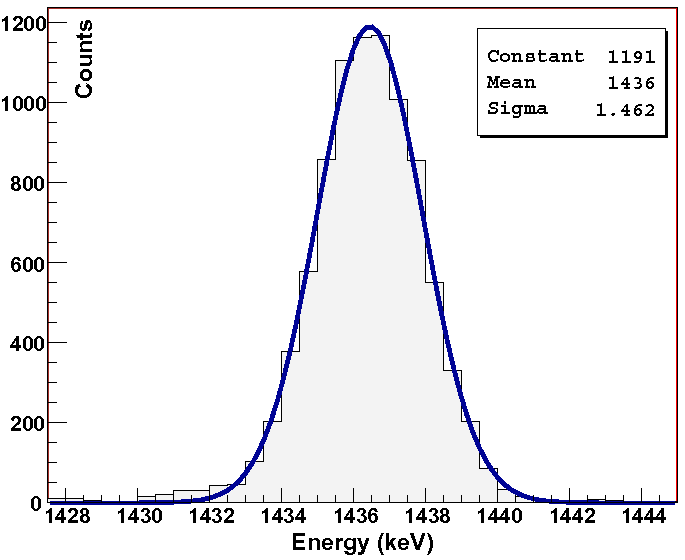}

\caption{\label{fig:Tracking-test-with}Spectra of tracked $\gamma$ rays emitted
in the reaction $\mathrm{d({}^{51}V,n)^{52}Cr}$ at $E_{\mathrm{LAB}}\left(^{51}\mathrm{V}\right)=77.0$
MeV. Left: The $\gamma$ rays were emitted at exactly $90^{\mathrm{o}}$
relative to the incoming beam and no Doppler correction was applied.
Right: The $\gamma$ rays were emitted isotropically into a $90^{\mathrm{o}}\pm20^{\mathrm{o}}$cone
covering the solid angle subtended by the triple cluster detector,
and a Doppler correction was applied. See text for further details.}

\end{figure}

Initially, two simulations were performed to verify that the tracking
and Doppler correction procedures functioned as expected. In the first
simulation, shown in the left panel of fig. \ref{fig:Tracking-test-with},
the 1434 keV $\gamma$ rays from the reaction $\mathrm{d\left(^{51}V,n\right)^{52}Cr}$,
at $E_{\mathrm{LAB}}\left(^{51}\mathrm{V}\right)=77.0$ MeV, were
emitted at an angle of exactly $90^{\mathrm{o}}$ relative to the
incoming beam. Gamma-ray tracking with a 5 mm smearing was used, but
no Doppler corrections were applied. In the second simulation, shown
in the right panel of fig. \ref{fig:Tracking-test-with}, the 1434
keV $\gamma$ rays were emitted isotropically into the $90^{\mathrm{o}}\pm20^{\mathrm{o}}$
cone covering the solid angle subtended by the triple cluster detector,
and both tracking (with 5 mm smearing) and Doppler corrections were
applied. The distance from the source to the front of the HPGe crystals
was 15.0 cm. The FWHM of the 1434 keV peak in both panels of fig.
\ref{fig:Tracking-test-with} are almost identical, which shows that
the $\gamma$-ray tracking and Doppler correction algorithms functioned
as expected.

Results of tracked and Doppler corrected $\gamma$-ray spectra from
simulations in which also the energy dependence of the intrinsic energy
resolution (see section \ref{sub:Experimental-detemination-of}) has
been included, are shown for the two reaction in fig. \ref{fig:Spectra-for-^{52}Cr}
and \ref{fig2:Photo-peak-for-^{38}Ar}. The smearing parameter was
kept at 5 mm and the $\gamma$ rays were emitted isotropically into
the $90^{\mathrm{o}}\pm20^{\mathrm{o}}$ cone.

\begin{figure}[H]
\begin{centering}
\includegraphics[scale=0.45]{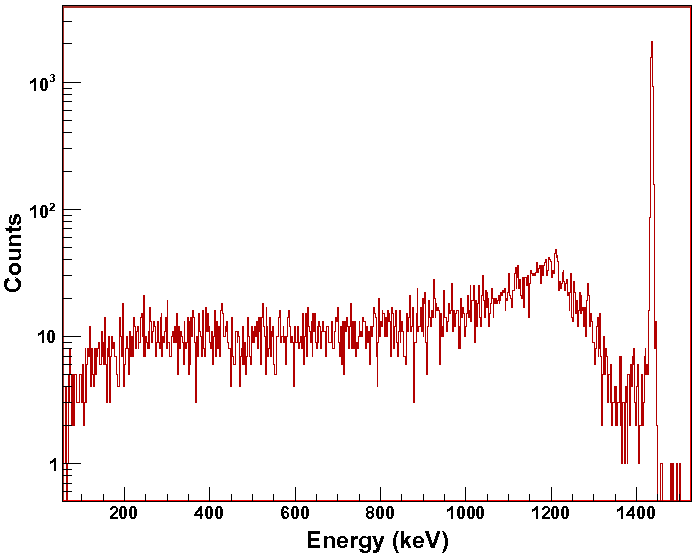}
\par\end{centering}

\caption{\label{fig:Spectra-for-^{52}Cr}Tracked and Doppler corrected $\gamma$-ray
spectrum of the 1434 keV $\gamma$ rays emitted in the reaction $\mathrm{d\left(^{51}V,n\right)^{52}Cr}$.
See text for further details. }

\end{figure}

\begin{table}
\begin{centering}
\begin{tabular}{|c|c|c|c|c|}
\hline 
Reaction & $E_{\gamma}$$\left[\mathrm{keV}\right]$ & $W\left[\mathrm{keV}\right]$ & $W_{\mathrm{i}}\left[\mathrm{keV}\right]$ & $W_{\mathrm{D}}\left[\mathrm{keV}\right]$\tabularnewline
\hline
\hline 
$\mathrm{d\left(^{37}Cl,n\right)^{38}Ar}$ & 2168 & 7.74 & 2.83 & 7.20\tabularnewline
\hline 
$\mathrm{d\left(^{51}V,n\right)^{52}Cr}$ & 1434 & 4.55 & 2.30 & 3.93\tabularnewline
\hline
\end{tabular}
\par\end{centering}

\caption{Contributions to the total FWHM, $W$, due to the intrinsic resolution,
$W_{\mathrm{i}}$, and the Doppler effects, $W_{\mathrm{D}}$, for
the two studied reactions.}

\end{table}

\begin{figure}[H]
\begin{centering}
\includegraphics[scale=0.79]{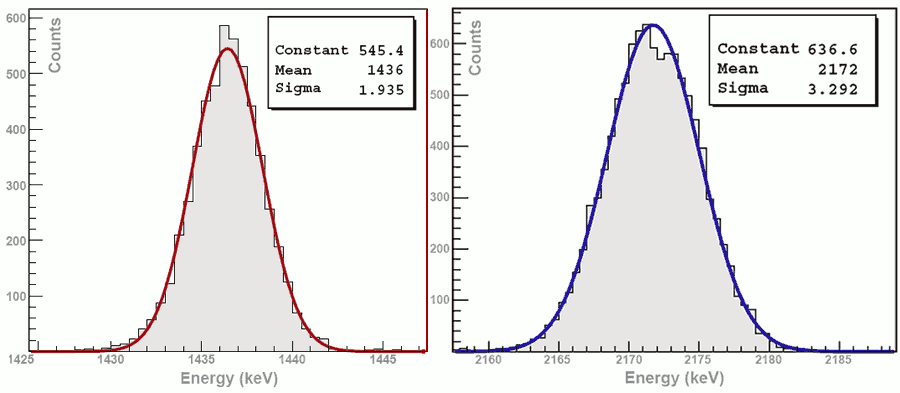}
\par\end{centering}

\caption{\label{fig2:Photo-peak-for-^{38}Ar}Tracked and Doppler corrected
$\gamma$-ray spectra of the 1434 keV (left) and 2168 keV (right)
$\gamma$ rays emitted in the reactions $\mathrm{d\mathrm{({}^{51}V,n)^{52}Cr}}$
($E_{\mathrm{LAB}}\left(^{51}\mathrm{V}\right)=77\:\mathrm{MeV}$)
and $\mathrm{d\mathrm{({}^{37}Cl,n)^{38}Ar}}$ ($E_{\mathrm{LAB}}\left(^{37}\mathrm{Cl}\right)=66.5\:\mathrm{MeV}$),
respectively. See text for further details.}

\end{figure}

Numerical values of the total FWHM $W$ of the peaks in fig. \ref{fig2:Photo-peak-for-^{38}Ar},
and their components (see eq. \ref{eq:FWHM}) due to the intrinsic
resolution ($W_{\mathrm{i}}$ obtained from fig. \ref{fig:Intrinsic-FHWM-as})
and due to the Doppler effects $W_{\mathrm{D}}$, are given in table
\ref{tab:Proton-and-Deuteron}. 

The results given in the table clearly show that the contribution
to the total FWHM, due to the Doppler effects, is much larger in the
reaction $\mathrm{d\left(^{37}Cl,n\right)^{38}Ar}$ than in the reaction
$\mathrm{d\left(^{51}V,n\right)^{52}Cr}$. The main reason for this
is the difference in mass number of the compound nuclei $^{38}\mathrm{Ar}$
and $^{52}\mathrm{Cr}$. This mass difference leads to a somewhat
larger and broader velocity distribution of the residual nuclei (see
fig \ref{fig:Velocity-distributions-for}) and, more importantly,
to an angular distribution of the residual nuclei which is both broader
and has a maximum at a larger angle (see right panel of fig. \ref{fig:angular effect doppler})
for the reaction $\mathrm{d\left(^{37}Cl,n\right)^{38}Ar}$ compared
to $\mathrm{d\left(^{51}V,n\right)^{52}Cr}$. The angular and velocity
distributions of the $\gamma$-ray emitting residual nuclei influence
directly the Doppler effects, which is seen as an effect on the FWHM
of the peaks.

\begin{figure}[H]
\begin{centering}
\includegraphics[scale=0.42]{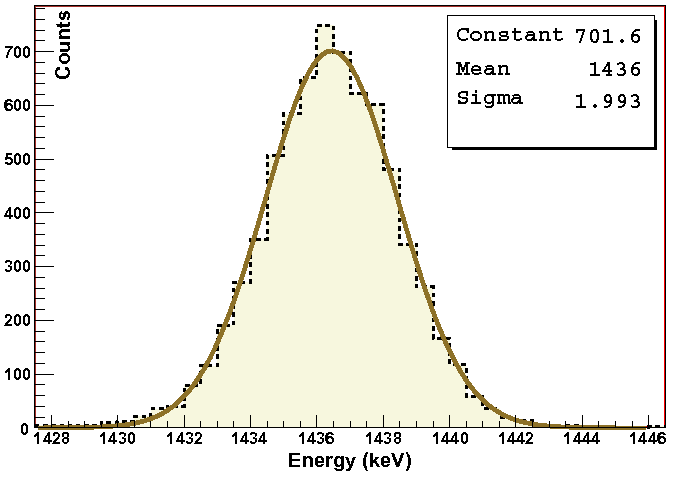}\includegraphics[scale=0.42]{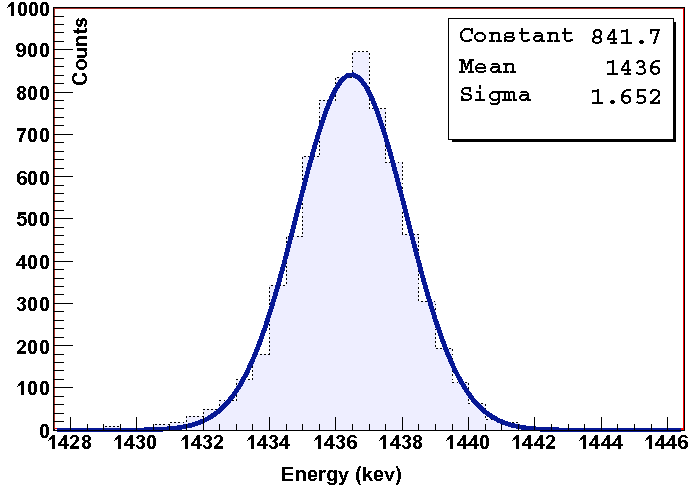}
\par\end{centering}

\caption{\label{fig:recoils}Comparison of the FWHM of the 1434 keV peak when
the Doppler correction was performed by using for the residual nucleus
an average velocity of $v/c=0.0548$ and an angle of $0^{\mathrm{o}}$
(left) and the exact value obtained event-by-event from evapOR (right).
The reaction was d$\mathrm{\left(^{51}V,n\right)^{52}Cr}$ at $E_{\mathrm{LAB}}\left(\mathrm{^{51}V}\right)=77.0$
MeV, the smearing parameter was 5 mm.}

\end{figure}

A simulation was also performed by correcting for the Doppler effects
introduced by the variation the velocity vector of the residual nucleus.
A comparison of the FWHM of the 1434 keV peak obtained when using
the average velocity vector and the precise value, as obtained event-by-event
from the evapOR simulation, is shown in fig \ref{fig:recoils}. For
the reaction $\mathrm{d\left(^{51}V,n\right)}$ at $E_{\mathrm{LAB}}\left(^{51}\mathrm{V}\right)=77.0$
MeV the FWHM is improved from 4.7 keV to 3.9 keV when the precise
value of the velocity vector of the residual nucleus is applied in
each event.

\subsection{Doppler effects induced by emission of $\alpha$ particles}

\begin{figure}[H]
\begin{centering}
\includegraphics[scale=0.47]{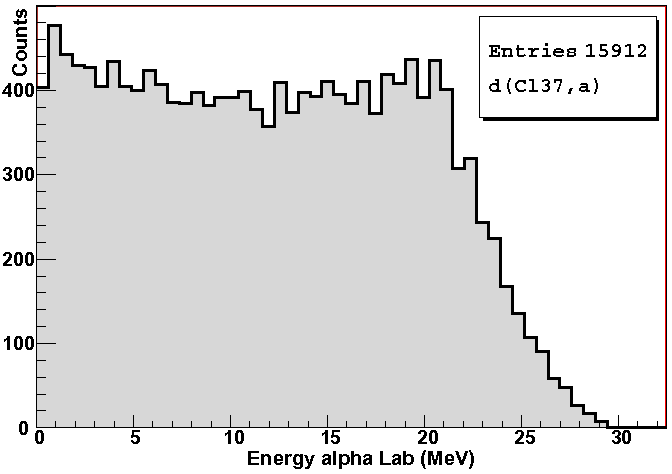}\includegraphics[scale=0.4]{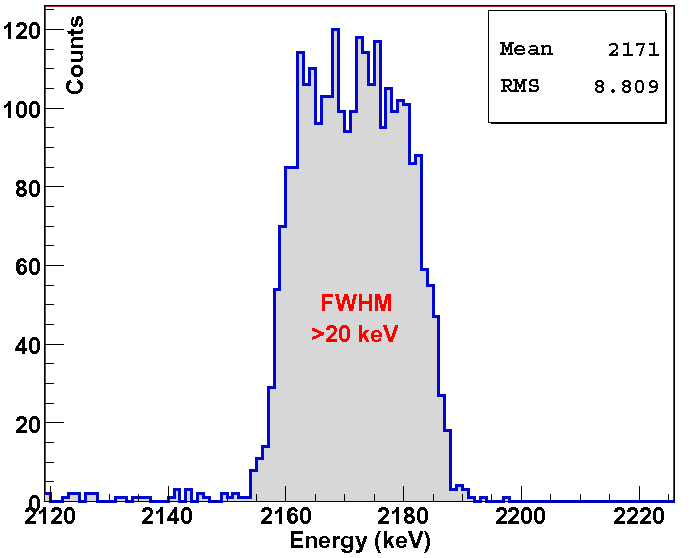}
\par\end{centering}

\centering{}\caption{\label{fig:a.-alpha}Doppler effects following the reaction $\mathrm{d}(^{37}\mathrm{Cl,}{\alpha})^{35}\mathrm{S}$
at $E_{\mathrm{LAB}}(^{37}\mathrm{Cl})=66.5$ MeV. Left: Kinetic energy
distribution of the residual nuclei $^{35}\mathrm{S}$. Right: FWHM
of a $\gamma$-ray peak at 2168 keV after tracking using the average
residual nucleus velocity vector for the Doppler correction . The
smearing parameter was 5 mm.}

\end{figure}

A simulation of Doppler effects, following a reaction channel in which
$\alpha$ particles are emitted, is shown in fig. \ref{fig:a.-alpha}.
The emitted $\alpha$ particles are four times heavier than neutrons
and have much higher kinetic energies in the laboratory system (compare
the left panels of fig. \ref{fig:angular effect doppler} and \ref{fig:a.-alpha}),
which leads to very large spreads both of the energy and angle of
the residual nuclei. This in turn leads to very broad $\gamma$-ray
peaks, as seen in the right panel of fig. \ref{fig:a.-alpha}. Thus,
for reaction channels with emission of $\alpha$ particles it is usually
necessary to determine event-by-event the velocity vector of the residual
nuclei.

\subsection{FWHM as a function of smearing parameter}

Simulations, in which the total FWHM of the 1434 keV and 2168 keV
peaks were determined as a function of the $\textsc{MGT}$ smearing
parameter, are shown in figures \ref{fig:fwhmvdPhoto-peak-for-^{38}Ar}
and \ref{fig:fqhmSpectra-for-^{38}Ar} for two different detector
distances, 15.0 cm and 23.5 cm, respectively. As seen in the figures,
the slopes increase for small values of the smearing parameter, up
to about 4-8 mm and become more or less constant for larger values.
The slopes are quite similar in all cases, although slightly larger
for the $\mathrm{d(^{51}V,n)^{52}Cr}$ reaction and for the shorter
distance of 15.0 cm. 

\begin{figure}[H]
\includegraphics[scale=0.44]{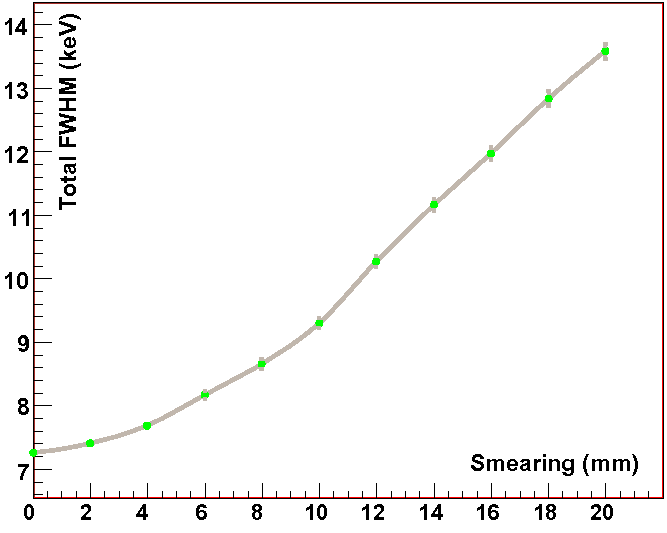}\includegraphics[scale=0.44]{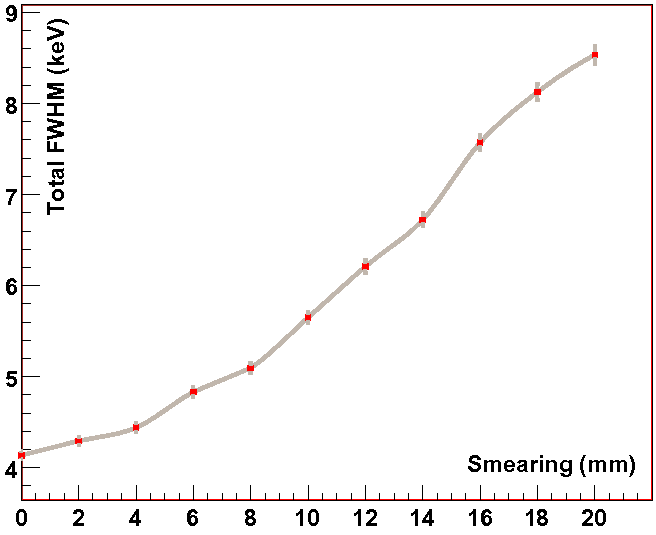}

\caption{\label{fig:fwhmvdPhoto-peak-for-^{38}Ar}FWHM as a function of the
MGT smearing parameter for the 2168 keV (left) and 1434 keV (right)
peaks of the $\mathrm{d(^{37}Cl,n)^{38}Ar}$ and $\mathrm{d(^{51}V,n)^{52}Cr}$
reactions, respectively. The distance from the source to the front
of the HPGe crystals was 15.0 cm. The data points are connected by
straight lines.}

\end{figure}

\begin{figure}[H]
\begin{centering}
\includegraphics[scale=0.39]{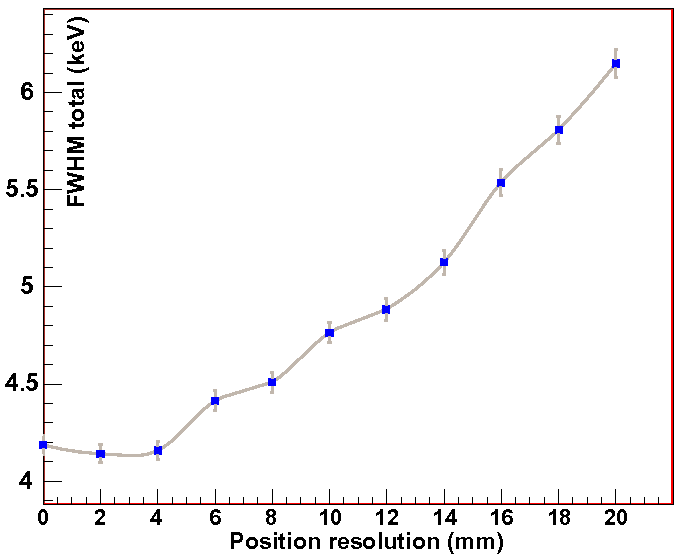}\includegraphics[scale=0.46]{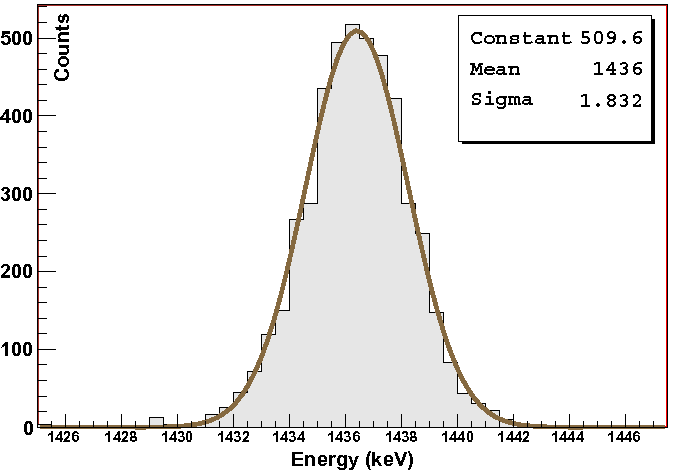}
\par\end{centering}

\caption{\label{fig:fqhmSpectra-for-^{38}Ar}Same as fig. \ref{fig:fwhmvdPhoto-peak-for-^{38}Ar}
but for a distance of 23.5 cm from the source to the front of the
HPGe crystals.}

\end{figure}

The interaction position resolution which is expected for the AGATA
HPGe crystals is of the order of 5 mm, a value which, however, is
not constant. The resolution will depend on the position of the interaction
in the crystal. For example in the front part of the crystal, where
the electric field is rather non-uniform, the interaction position
resolution will be worse compared to a position in the central parts.
The interaction position resolution will also depend on the energy
deposited in the interaction point. It is expected that it will be
proportional to the inverse square root of the interaction energy,
a dependence which is implemented in MGT, although not used in this
work.

A conclusion of the results shown in fig. \ref{fig:fwhmvdPhoto-peak-for-^{38}Ar}
and \ref{fig:fqhmSpectra-for-^{38}Ar} is that with the proposed reactions,
$\mathrm{d(^{37}Cl,n)^{38}Ar}$ at $E_{\mathrm{LAB}}(^{37}\mathrm{Cl})=66.5$
MeV and $\mathrm{d(^{51}V,n)^{52}Cr}$ at $E_{\mathrm{LAB}}(^{51}\mathrm{V})=77.0$
MeV, it will be difficult to get a good determination of the interaction
position resolution from the measured FWHM of the peaks. One possibility
to improve the sensitive, is to make the measurements for much smaller
distances, than what was used in this work (15.0 cm and 23.5 cm).
Another possibility is to try to find another reaction with larger
average recoil velocities but which still has small enough spread
of the residual nucleus velocity vector so that the average Doppler
correction procedure may be used.

\section{Conclusions and summary }

In section \ref{sec:TALYS} the studied reactions in TALYS resulted
in two interesting reactions for further study, namely $\mathrm{d({}^{51}V,n)^{52}Cr}$
and $\mathrm{d({}^{37}Cl,n){}^{38}Ar}$. On the basis of the requirements
in section \ref{sub:Requirements}, these reactions have suitable
$\gamma$-ray energies and life times within the required range. Both
reactions have dominant cross sections for the desired neutron channel.
Both reactions lead to even-even residual nuclei and give quite large
recoil velocities (important in order to study the Doppler effects). 

In section \ref{sec:Results-target-studies}, the target effects were
investigated using TRIM. Two major sources of unwanted target effects
were studied and determined to be negligible for the Doppler shifts.
The average energy loss for the ions in the target material were simulated
and presented.

In section \ref{sec:evapOR}, results from evapOR and TALYS were compared.
Different energy and angular distributions were produced by evapOR.
Due to the different amount of nucleons in the compound nuclei of
the two reactions, the $^{38}\mathrm{Ar}$ residual nuclei were deflected
more from the ion beam direction compared to $^{52}\mathrm{Cr}$.
This difference in angular deflection is the reason why the Doppler
broadening is larger for the reaction $\mathrm{d({}^{37}Cl,n){}^{38}Ar}$
than for $\mathrm{d({}^{51}V,n)^{52}Cr}$. 

In section \ref{sec:Spectra} the detection of the $\gamma$ rays
in one AGATA triple cluster detectors was simulated with \noun{Geant4}.
The $\gamma$ rays were tracked by the MGT tracking program to produce
Doppler corrected $\gamma$-ray spectra. The FWHM of the $\gamma$
ray peaks of interest were calculated as a function of the interaction
position resolution. The $\mathrm{d({}^{51}V,n)^{52}Cr}$ reaction
was found to be slightly more suitable than the $\mathrm{d\left(^{37}Cl,n\right)^{38}Ar}$
reaction for the planned AGATA commissioning experiment. 

\appendix

\chapter{Acknowledgment}

\begin{onehalfspace}
This diploma work could not have been fulfilled without the great
support, aid and guidance from two excellent supervisors. Professor
Johan Nyberg and the PhD student, P\"{a}r-Anders S\"{o}derstr\"{o}m. Johan shared
his knowledge generously and delivered me a comprehensive introduction
into the field of nuclear physics research. His supervising showed
wide patience and broad capacity of describing the tasks and the methods
of solving them. P\"{a}r-Anders, I am obliged to you, for all hours with
programming support. All our general and specific discussion about
physics, truly enlightened my mind. It was important for me to share
that room with you in order to collect as much as possible of your
experience. I would like to thank all the people at the division of
nuclear and particle physics, it was a pleasure to write the diploma
work among you people. Thanks for my opponent Pelle who gave my a
proper review of my report. Special thanks for Mikael, Emma, Bengt,
Patrik, Carl-Oscar, Samson, Vasily, Pernilla, Riccardo, Henrik and
Henrik. I want also to pay my respect for professor Jan Blomgren who
supported and guided during my application for the PhD position. 

My deepest gratitude goes to my dear family. You are always present,
no matter the circumstances. This diploma work is dedicated to my
fantastic parents, who I stand in obligation to for my upbringing.
You supported me in my choices through life, that meant everything
for me. My siblings, Amir and Afnan, you are making me proud. My dear
Sheima, thanks for painting my entire world. 

My friends, you have your part in this as well. Greetings to Alaa,
Alvaro, Martin, PO, Rickard, Jonatan, Fabrice, Chi, Ludde, Mortada,
Ali, Niklas, Jan, Tassilo, Miguel, Hans-Erik, Pedro, Zeyd and all
you others.
\end{onehalfspace}

\chapter{\label{cha:Available-ion-beams}Available ion beams at Laboratori
Nazionali di Legnaro (Padova) }

\begin{table}[H]
\caption{Available ion beams at LNL}

\centering{}\begin{tabular}{|c||c||c||c||c||c|}
\hline 
\textbf{Beam} & \textbf{Current in (nA)} & \textbf{Beam} & \textbf{Current in (nA)} & \textbf{Beam} & \textbf{Current in (nA)}\tabularnewline
\hline
\hline 
\textcolor{blue}{1H } & \textit{1000} & \textcolor{blue}{40Ca} & \textit{150} & \textcolor{blue}{79Br} & \textit{800}\tabularnewline
\hline 
\textcolor{blue}{2H} & \textit{300} & \textcolor{blue}{48Ca} & \textit{150} & \textcolor{blue}{81Br} & \textit{800}\tabularnewline
\hline 
\textcolor{blue}{6Li} & \textit{50} & \textcolor{blue}{48Ti} & \textit{700} & \textcolor{blue}{90Zr} & \textit{100}\tabularnewline
\hline 
\textcolor{blue}{7Li} & \textit{200} & \textcolor{blue}{50Cr} & \textit{150} & \textcolor{blue}{91Zr} & \textit{25}\tabularnewline
\hline 
\textcolor{blue}{10B} & \textit{100} & \textcolor{blue}{52Cr } & \textit{150 } & \textcolor{blue}{92Zr} & \textit{30}\tabularnewline
\hline 
\textcolor{blue}{11B} & \textit{400} & \textcolor{blue}{51V} & \textit{250} & \textcolor{blue}{94Zr} & \textit{30}\tabularnewline
\hline 
\textcolor{blue}{12C} & \textit{3000} & \textcolor{blue}{54Fe} & \textit{200} & \textcolor{blue}{96Zr} & \textit{300}\tabularnewline
\hline 
\textcolor{blue}{13C} & \textit{35} & \textcolor{blue}{56Fe} & \textit{200} & \textcolor{blue}{92Mo} & \textit{250}\tabularnewline
\hline 
\textcolor{blue}{14N } & \textit{800} & \textcolor{blue}{58Ni} & \textit{1500} & \textcolor{blue}{94Mo} & \textit{150 }\tabularnewline
\hline 
\textcolor{blue}{16O} & \textit{2000} & \textcolor{blue}{60Ni} & \textit{500} & \textcolor{blue}{95Mo} & \textit{250}\tabularnewline
\hline 
\textcolor{blue}{17O} & \textit{500} & \textcolor{blue}{64Ni} & \textit{500} & \textcolor{blue}{96Mo} & \textit{250}\tabularnewline
\hline 
\textcolor{blue}{18O} & \textit{500} & \textcolor{blue}{63Cu} & \textit{1000} & \textcolor{blue}{97Mo} & \textit{150}\tabularnewline
\hline 
\textcolor{blue}{19F} & \textit{3000} & \textcolor{blue}{65Cu} & \textit{350} & \textcolor{blue}{98Mo} & \textit{400}\tabularnewline
\hline 
\textcolor{blue}{24Mg} & \textit{300 } & \textcolor{blue}{64Zn} & \textit{500} & \textcolor{blue}{100Mo} & \textit{150}\tabularnewline
\hline 
\textcolor{blue}{26Mg} & \textit{200} & \textcolor{blue}{66Zn} & \textit{250} & \textcolor{blue}{96Ru} & \textit{200}\tabularnewline
\hline 
\textcolor{blue}{27Al} & \textit{400} & \textcolor{blue}{68Zn} & \textit{200} & \textcolor{blue}{98Ru} & \textit{70}\tabularnewline
\hline 
\textcolor{blue}{28Si} & \textit{1000} & \textcolor{blue}{69Ga} & \textit{300} & \textcolor{blue}{99Ru} & \textit{450}\tabularnewline
\hline 
\textcolor{blue}{29Si} & \textit{300} & \textcolor{blue}{70Zn} & \textit{200} & \textcolor{blue}{100Ru} & \textit{450}\tabularnewline
\hline 
\textcolor{blue}{30Si} & \textit{200} & \textcolor{blue}{71Ga} & \textit{200} & \textcolor{blue}{101Ru} & \textit{600}\tabularnewline
\hline 
\textcolor{blue}{31P} & \textit{500} & \textcolor{blue}{74Ge} & \textit{800} & \textcolor{blue}{102Ru} & \textit{1000}\tabularnewline
\hline 
\textcolor{blue}{32S} & \textit{2000} & \textcolor{blue}{76Ge} & \textit{200} & \textcolor{blue}{104Ru} & \textit{650}\tabularnewline
\hline 
\textcolor{blue}{33S} & \textit{400} & \textcolor{blue}{76Se} & \textit{200} & \textcolor{blue}{107Ag} & \textit{400}\tabularnewline
\hline 
\textcolor{blue}{34S} & \textit{200 } & \textcolor{blue}{77Se} & \textit{300} & \textcolor{blue}{109Ag} & \textit{400}\tabularnewline
\hline 
\textcolor{blue}{36S} & \textit{400} & \textcolor{blue}{78Se} & \textit{1000} & \textcolor{blue}{127I} & \textit{800}\tabularnewline
\hline 
\textcolor{blue}{35Cl} & \textit{2000} & \textcolor{blue}{80Se} & \textit{2000} & \textcolor{blue}{197Au} & \textit{500}\tabularnewline
\hline 
\textcolor{blue}{37Cl} & \textit{650} & \textcolor{blue}{82Se} & \textit{400} &  & \tabularnewline
\hline
\end{tabular}
\end{table}

\chapter{Simulation steps}

\subsubsection*{1. evapOR }
\begin{itemize}
\item Run evapOR. 
\item Modify the {*}.inp file for evapOR, where you can choose nuclei, spin
and energy. 
\item evapOR-loop 
\item evapOR-loop {*}.inp 
\end{itemize}
You will have now some produced files. The important one is {*}.pax

\subsubsection*{2. Sortpax: }

You run sortpax to generate the Doppler shifted gammas and to produce
the {*}.aga file. 
\begin{itemize}
\item sortpax {*}.pax 
\item choose aga file with typing '{*}.aga' 
\item choose txt file to read the events, with typing '{*}.txt' 
\item choose bin file for evapOR Root tree, with typing '{*}.bin' 
\item choose the gate on nucleus (Z, N) 
\item choose the angle of emitted photons (thetamin, thetamax, phimin, phimax) 
\end{itemize}
Now you will have the {*}.aga file. (This you copy to fari in my case)

\subsubsection*{3. AGATA \noun{Geant4}: }

Go to directory of AGATA \noun{Geant4} 
\begin{itemize}
\item Run 'Agata -Ext -n' in order to read the particles also. 
\item Type the desired of the following commands

\begin{itemize}
\item /Agata/file/enableLM 
\item /Agata/file/verbose 1 to write information 
\item /Agata/file/packingDistance 5. for packing 
\end{itemize}
\item Define the geometry, a cluster is given by this macro 

\begin{itemize}
\item /control/execute macros/geom180Ali.mac 
\item Look in macro file and manual agata \noun{Geant4} to see the other
geometries. The distance to the detector is given in A180eulerAli.list
which is under the folder A180. You can there put and move the detector
anywhere. 
\end{itemize}
\end{itemize}
To generate with the {*}.aga file use 
\begin{itemize}
\item /Agata/generator/emitter/eventFile /home/tsl0/ali/Desktop/doppler.aga
/Agata/run/beamOn 100000 
\end{itemize}
To generate with AGATA 
\begin{itemize}
\item /Agata/generator/recoil/beta 5. (for nucleu's velocity)
\item /Agata/generator/gamma/energy 1000 (the energy of gammas)
\item /Agata/generator/gamma/thetaRange 65.0 115.0 (the $\theta$ angle
emitted gammas) 
\item /Agata/generator/gamma/phiRange 65.0 115.0 (the $\phi$ angle emitted
gammas)
\end{itemize}

\subsubsection*{4. MGT:}

The output is gammaevents.0000/1/2..... and this should be used in
MGT go the folder where this file is and type 
\begin{itemize}
\item ' mgt -fw 0 0 -f GammaEvents.0000 -ll 0 -oa 30 30 -dd 5 5 -sr 0.05
0 0 '
\end{itemize}
where fw is defining the intrinsic FWHM, statistical and then noise
factor. ll 0/1/2/3 gives more information. dd defines the smearing
and packing . 
\begin{itemize}
\item sr gives the average velocity and direction in $\theta$ and $\phi$.
\end{itemize}

\end{document}